\documentclass[12pt,letterpaper,a4paper]{article}
\usepackage[includeheadfoot,
            marginratio={1:1,2:3},
            width=412pt,
            height=688pt,]{geometry}
\usepackage{amsmath}
\usepackage{amsfonts}
\usepackage{amssymb}
\usepackage{graphicx}
\usepackage{empheq}
\usepackage{cite}
\usepackage{color}
\usepackage{hyperref}

\usepackage{float}
\restylefloat{table}
\restylefloat{figure}

\newcommand{\nc}{\newcommand}
\nc{\beq}{\begin{equation}}
\nc{\eeq}{\end{equation}}
\nc{\bea}{\begin{eqnarray}}
\nc{\eea}{\end{eqnarray}}
\newcommand{\eq}[1]{\begin{equation}
                     \begin{split} #1 \end{split}
                     \end{equation}}
\def\ov{\overline}

\newdimen\csize\csize=1.5ex
\def\young#1{\tiny\vcenter{\hbox{\vrule\vtop{\hrule
  \offinterlineskip\halign{&\vbox
  {\hbox to\csize {\strut\hss##\hss\vrule}\hrule}\cr#1 \crcr}}}}}

\begin{document}

\vspace*{-1.5cm}
\begin{flushright}
  {\small
  MPP-2013-184\\
  }
\end{flushright}

\vspace{1.5cm}
\begin{center}
  {\Large  F-term Stabilization of Odd Axions in \\ LARGE Volume Scenario}

\end{center}

\vspace{0.75cm}
\begin{center}
Xin Gao$^{\dagger, \,  \ddagger}$\footnote{Email: gaoxin@mppmu.mpg.de} and Pramod Shukla$^\dagger$\footnote{Email: shukla@mppmu.mpg.de}
\end{center}

\vspace{0.1cm}
\begin{center}
\emph{
$^{\dagger}$ Max-Planck-Institut f\"ur Physik (Werner-Heisenberg-Institut), \\
   F\"ohringer Ring 6,  80805 M\"unchen, Germany\\
\vskip 1cm
$^{\ddagger}$ State Key Laboratory of Theoretical Physics, \\
Institute of Theoretical Physics,\\ Chinese Academy of Sciences, P.O.Box 2735, Beijing 100190, China } \\[0.1cm]
\vspace{0.2cm}

 \vspace{0.5cm}
\end{center}

\vspace{1cm}

%%%%%%%%%%%%%%%%%%%%%%%%%%%%%%%%%%%%%%%%%%%%%%%%%%%%%%%%%%%%%%%%%%%%%%%%%%%%%%%%%%%%%%%%%%%%%

\begin{abstract}
In the context of the LARGE volume scenario,
stabilization of  axionic moduli is revisited.
This includes both  even and odd axions with
their scalar potential being generated by F-term contributions via various
tree-level and non-perturbative effects like
fluxed E3-brane instantons and fluxed poly-instantons.
%In the toy model setup, we investigate the possibility to
%stabilize odd moduli only by $F$-term.
In all the cases, we estimate the decay constants and masses
of the axions involved.
%, thus  aiming for
%candidates of axions that might be relevant at  low energies
%like the  QCD axion.
\end{abstract}

\clearpage

%%%%%%%%%%%%%%%%%%%%%%%%%%%%%%%%%%%%%%%%%%%%%%%%%%%%%%%%%%%%%%%%%%%%%%%%%%%%%%%%%%%%%%%%%%%%%%

\tableofcontents
\section{Introduction}
\label{sec:intro}

From the point of view of constructing (semi-)realistic models in string compactifications, the understanding of moduli stabilization is a very central issue and has been under deep investigation for more than a decade. In order to quantitatively address certain aspects of cosmology and of particle physics, moduli stabilization is a prerequisite, as on the one hand some physical parameters depend on the value of the moduli and on the other hand the existence of such massless scalars is incompatible with observations.

The standard paradigm for string moduli stabilization is described (and
better understood) in the framework of type IIB orientifolds with
$O7$ and $O3$-planes, and two popular classes of models, namely KKLT \cite{Kachru:2003aw} and LARGE volume scenarios \cite{Balasubramanian:2005zx}, have been in the market for almost a decade. In these models, a combination of background
three-form fluxes and D3-brane instantons can lead to a potential
for the axion-dilaton, the complex structure and the K\"ahler moduli
\cite{Gukov:1999ya,Becker:2002nn,Grana:2005jc,Lust:2006zg}. Usually, the  moduli stabilization scheme in these two classes of models is a two-step procedure. In first step, one stabilizes complex structure moduli along with axion-dilaton at the leading order via a tree level K\"ahler potential and a perturbative flux-contribution to the superpotential. The K\"ahler moduli which remains flat (due to a so-called  `no-scale structure') at this stage are lifted in a second step by including the sub-leading non-perturbative corrections to the superpotential $W$, and the same results in a supersymmetric (KKLT) $AdS$-minimum \cite{Kachru:2003aw}.
Taking also the leading order perturbative $\alpha^\prime$-corrections to the
K\"ahler potential into account, a non-supersymmetric $AdS$-minimum
at large overall volume appears. This is the so-called
LARGE volume scenario (LVS) \cite{Balasubramanian:2005zx}, which has been exploited  in
the literature in the context of getting
realistic particle physics and realizing inflationary cosmology both
(See \cite{Conlon:2005jm,Conlon:2007gk} and references therein).
Recently,  the orientifold even axionic  sector was scrutinized
 \cite{Cicoli:2012sz} leading to the proposal
that in the context  of the LVS
there exists  a whole axiverse,
which means that the decay constants of the different axions
vary over a wide range of values. 
This is mainly owed to the
fact that different axions get different volume suppression
factors in their kinetic terms. 
%However, this analysis is only to take account the decay constant 
%by ensuring axions to be massless . 
Further, most of these studies are only focused on the orientifold even sector of axions.
A detailed analysis in these directions with orientifold odd axionic sector is missing. However, in some models, moduli stabilization \cite{Hristov:2008if}, inflationary aspects \cite{Grimm:2007hs,McAllister:2008hb,Flauger:2009ab} as well as particle pheno model building \cite{Blumenhagen:2008zz} with the inclusion of involutively odd (1,1)-cohomology sector have been initiated in the meantime.
%However, if the axion
%also appear in the flux/instanton induced scalar potential,
%they generically are stabilized at the same scale as
%the corresponding saxion, which due to the above mentioned
%cosmological constraints
%must be  much heavier than for instance the mass of the QCD axion.
%Therefore, the QCD axion as well as
%any other low-energy axion-like particle (ALP)
%should better stay nearly massless after moduli stabilization.
%Realizing all these details in a concrete string compactification
%is a challenging problem that  becomes even more acute by
%the need to actually couple a candidate axion to QCD
%in the standard way without interfering with the assumptions
%already made in the analysis of moduli stabilization.
%Here we have in mind  the string constraints governing  the
%coexistence of D7-branes, fluxes and instantons, like
%the chirality problem pointed out in \cite{Blumenhagen:2007sm} or the
%Freed-Witten anomalies \cite{Freed:1999vc}.

These orientifold odd axion also play a crucial role in global model building in
string compactification.
Here we have in mind  the string constraints governing  the
coexistence of D7-branes, fluxes and instantons, like
the chirality problem pointed out in \cite{Blumenhagen:2007sm} or the
Freed-Witten anomalies \cite{Freed:1999vc}.
The chirality issue comes %when one considers a visible sector in a compactification scheme, there are
with the appearance of extra  zero modes located at the intersection between the instanton E3-brane and D7-brane supporting the
visible sector.  This
prevents a class of instantons from participating in moduli stabilization.
Several approaches have been proposed in building up models which (could) avoid such a problem.
One way is, not to support the visible sector on 
the divisor which gives rise to the non-perturbative superpotential contribution \cite{Collinucci:2008sq,Cicoli:2011qg,
Balasubramanian:2012wd,Cicoli:2012vw}. 
Models which support visible sector with D-branes at singularities have been proposed in
\cite{Balasubramanian:2012wd,Cicoli:2012vw} in which
one needs to embed such singularities in a compact Calabi-Yau threefold $X$
with non-zero odd components in cohomology class $H^{1,1}_-(X/\sigma)$. 
%under some holomorphic involution $\sigma$.
Another way to avoid the chirality issue is to
include the gauge flux on the instanton E3-brane supported by the orientifold odd two-cycle \cite{Grimm:2011dj}. 

For having the fluxed-instanton contribution, one needs the involutively odd-moduli ($b^a, c^a$) which 
arise from the NS-NS field $B_2$ and R-R field $C_2$ in type IIB orientifolds to correct the E3-brane
superpotential and remove the extra charged zero modes. 
%These $b^a$ and $c^a$ moduli appear when there is a non-trivial splitting of cohomology $H^{1,1}(X)$ 
%under holomorphic involution $\sigma$, and  are counted by  $h^{1,1}_-(X)\ne 0$. 
%In order to have fluxed instanton contributions,  % we generalize  the analysis initiated  in \cite{Cicoli:2012sz}
%in two directions. First, 
%we need to consider pure axionic chiral multiplets for which
%the entire complex boson is made from axions. These can appear
%in Type IIB orientifolds if $h^{1,1}_-(X)\ne 0$. In this case,
%the dimensional reduction of the two ten-dimensional axions $B_2$ and
%$C_2$ on such an odd two-cycle do survive the orientifold projection.
These odd axions combine in pure axionic chiral multiplets for which
the entire complex boson is made from axions. These new chiral multiplets $G^a$ appear in the effective action, i.e.
in the K\"ahler- and superpotential in a completely different
manner than the even moduli so that they must be treated
separately. 
In addition,
we also include those axions
sitting in the same chiral multiplet as the saxions
governing the size of four-cycles having the right topology
to support  so-called poly-instantons.
These are  sub-leading non-perturbative contributions which can be briefly
described as instanton corrections to instanton actions.
These were introduced
in  \cite{Blumenhagen:2008ji}, further elaborated on in
\cite{Petersson:2010qu} and have been analyzed
recently in the context of the LVS in 
\cite{Blumenhagen:2008kq,Cicoli:2011yy,Cicoli:2011ct,Cicoli:2012cy,Blumenhagen:2012Poly1,
Cicoli:2012tz,Blumenhagen:2012ue,Gao:2013hn, Lust:2013kt}.
In order to support the odd moduli in  models of (type IIB) string compactification, 
a classification of the involutions on the toric Calabi-Yau threefolds with $h^{1,1}\le4$ which may result
 non-zero  odd component of (1,1)-cohomology class is also studied in \cite{Gao:2013pra}.
%In \cite{Blumenhagen:2012Poly1}, we have clarified the zero mode conditions for and Euclidean
%D3-brane instanton, wrapping a so-called ``wilson" divisor of the threefold,
%to generate such a poly-instanton effect.
%There we have also provided the construction of some concrete models having the right
%divisors to both feature the LARGE volume scenario and to support additional poly-instanton corrections.
%Building on the new result, the  consequently moduli stabilization and single-filed inflation
%is studied \cite{Blumenhagen:2012ue}.
%Different from the usual K\"ahler inflation(blow-up model)
%and fiber inflation, here the inflaton is neither a blow-up mode nor the size modulus of the K3 fiber but actually a K\"ahler
%modulus corresponding to a so-called "Wilson" line volume divisor.
%More recently, we generalize the poly-instanton inflation driven by such wilson divisor volume modulus
%to include the respective axion modulus. This two-field dynamics results in a ``Roulette" type inflation,
%where we find the large local non-Gaussianities can be generated in the beyond slow-roll
%regime \cite{Gao:2013hn}.
%The possibility of generating primordial non-Gaussianities in the slow-roll as well as in the beyond slow-roll region is investigated.
%We find that although the non-linearity parameters are quite small during the slow-roll regime,
%the same are significantly enhanced in the beyond slow-roll regime \cite{}.

In general, axions play an important role in physics beyond the Standard Model
and, via axion driven inflation, might provide a
bridge between particle physics with  cosmology. 
On the particle physics side,
axions were introduced to solve the
strong CP problem of QCD \cite{Weinberg:1977ma,Peccei:1977hh,Wilczek:1977pj,Kim:1986ax,Turner:1989vc,Kim:2008hd}. 
%From the non-observation of
%the electric dipole moment of the neutron, for the  CP-violating $\theta$-term
%one derives the strong constraint  $|\theta| < 10^{-11}$.
%For QCD axion, from the energy loss of astrophysical objects one gets
%the lower bound $f_a> 10^9\,$GeV.
%From the axion contribution to the dark matter energy density of the
%universe one can also derive an upper bound, which however
%depends on the so-called initial misalignment angle.
%In the most  natural case, this leads to $f_a<10^{12}\,$GeV, but
%can be larger by allowing  some tuning.
%For realizing inflation the essential ingredient is a (pseudo-)scalar field, which has a sufficiently flat potential allowing for a long period of slow roll. 
However, due to the shift symmetry of the axions, a non-trivial  subleading potential
is generated at the non-perturbative level, making
them ideal inflaton candidates as well %For consistency with the recent observations of anisotropies in the cosmic microwave background(CMB), the axion has to roll over trans-Planckian distances, i.e. these models  are of large field inflation type with $f_{\rm{infl}}\gtrsim  M_{\rm pl}$. This  leads to a  mass of the  inflationary axion around the GUT or string scale
\cite{Kim:1986ax,Turner:1989vc,Freese:1990rb,Sikivie:2006ni,Raffelt:1996wa,Komatsu:2008hk}\footnote{See  \cite{Kim:2002tq} for applications of axions as quintessence fields, and\cite{Chatzistavrakidis:2012bb} for an interesting field theoretic attempt of combining the three axionic scenarios (QCD axion, inflaton and quintessence axion) into a single framework.}. Since in most of the axionic model building purposes (e.g. in axionic inflation),  the axionic decay constants  are required to be very high,
a treatment in a UV-complete theoretical framework such
as string theory is desirable and has indeed been pursued.
Axionic fields are already ubiquitous in superstring theories
in ten-dimensions and via compactification
generically lead to the order of $10-10^2$ axionic fields
in four dimensions \cite{Arvanitaki:2009fg,Arvanitaki:2010sy}. 
%Their dynamics is quite intricate as
%axions play an important role in the Green-Schwarz mechanisms,
%where they mix with the abelian gauge fields in order to cancel
%the gauge anomalies. Axions often appear in a chiral supermultiplet,
%where they are combined with a scalar field, a so-called saxion,
%which describes the deformation of the underlying compact geometry.
%These geometric moduli can  for instance be the
%K\"ahler or complex structure moduli
%of Calabi-Yau manifolds. They  need to be
%stabilized for a  string model to be realistic, as
%they lead to unobserved fifth forces and interfere with
%the standard big bang cosmology, in particular
%big bang nucleosynthesis \cite{RevModPhys.29.547,Arvanitaki:2009fg}.
These axions often appear in a chiral supermultiplet,
where they are combined with a scalar field, a so-called saxion,
which describes the deformation of the underlying compact geometry.
These geometric moduli can  for instance be the
K\"ahler or complex structure moduli
of Calabi-Yau manifolds. For a  string model to be realistic, all moduli have to be
stabilized as
they lead to unobserved fifth force and interfere with
the standard big bang cosmology, in particular
big bang nucleosynthesis \cite{Arvanitaki:2009fg,Arvanitaki:2010sy,RevModPhys.29.547}.

{\it In this article, our main focus is to revisit the F-term moduli stabilization in LARGE volume scenarios. The idea is to include the involutively odd axions and instanton flux effects to generate F-term contributions depending on odd axions. Addressing more involved issues, for example, the ones mentioned in the aforementioned paragraph need more concreteness in the setup, and are beyond the scope of this article}. To be specific, we limit ourselves to a toy-model setup, and without supporting a concrete MSSM-like visible sector on D7-brane wrapping a holomorphic divisor, we assume that concrete setups with desired divisor intersections and allowed fluxes consistent with tadpole/anomaly cancellations can be constructed. Apart from moduli stabilization, as a by-product of our investigations regarding the simple estimates of volume scalings in axion decay constants and masses, we explore the possible mass hierarchy among the various moduli.

The article is organized as follows. In section \ref{sec_Setup}, we start with
a brief review of a generic Type IIB orientifold framework, and
following \cite{Grimm:2011dj}, we collect
the relevant ingredients on fluxed $D3$-brane instanton contributions to the
superpotential.
In section \ref{sec_Even-axions},
we discuss the moduli stabilization in an extended LARGE volume setup with
the inclusion of a single involutively odd axion and fluxed-instanton effects.
In section \ref{sec-PolyLVS},
we extend the analysis for a fluxed poly-instanton setup. For all the cases, we also present some estimates for the decay constants and masses of various even/odd axions.
%which could be relevant for looking at the possibility of realizing a QCD axion candidate.
%Special emphasis is here on finding correlations
%of the mass scale of the axion to the mass scale of their
%partner saxions and on axions which stay ultra-light after
%moduli stabilization. 
Section \ref{sec_Conclusions} presents the overall conclusions followed by a short appendix of the relevant intermediate expressions.

% we start with a short review of the moduli stabilization mechanism after the
%implementing the poly-instanton corrections in the standard LARGE volume scenario.
% we analyze two classes of even axionic  fields.
%To get started, we revisit the setup of a
%standard swiss-cheese LVS setup and consider
%the axions related to  the big and the small divisor.
%This case was already discussed in \cite{Cicoli:2012sz}.
%Second,  we include in our analysis axions arising  from
%four-cycles with the topology of a Wilson line divisor that
% can generate a poly-instanton correction.
%In section \ref{sec_Odd-axions}, we consider LVS moduli stabilization
%with the inclusion of the orientifold odd axions.
%In all cases, we  compute the decay constants and masses of  all the
%axions involved. Special emphasis is here on finding correlations
%of the mass scale of the axion to the mass scale of their
%partner saxions and on axions which stay ultra-light after
%moduli stabilization.

\section{Preliminaries}
\label{sec_Setup}

%In this section, we review some basics of type IIB orientifolds 
%in the LARGE volume scenario.

%\subsection{Type IIB orientifolds with fluxes and instantons}

Let us review some of the basic ingredients in Type IIB orientifold
compactifications with $O7$ and $O3$-planes. Here we focus on those
aspects which will become relevant in our investigation
of axions in the LARGE volume scenario.

\subsubsection*{Type IIB Orientifolds}

We consider Type IIB superstring theory compactified on an orientifold of a
Calabi-Yau threefold $X$.
The admissible orientifold projections fall into two categories,
which are distinguished by their action on the
K\"ahler form $J$ and the holomorphic three-form $\Omega_3$ of
the Calabi-Yau:
\begin{eqnarray}
\label{eq:orientifold}
 {\cal O}= \begin{cases}
                       \Omega_p\, \sigma \qquad &{\rm with} \quad
                       \sigma^*(J)=J\,,\quad  \sigma^*(\Omega_3)=\Omega_3 \, ,\\[0.1cm]
                       (-)^{F_L}\,\Omega_p\, \sigma\qquad & {\rm with} \quad
        \sigma^*(J)=J\,, \quad \sigma^*(\Omega_3)=-\Omega_3
\end{cases}
\end{eqnarray}
where $\Omega_p$ is the world-sheet parity transformation and $F_L$ denotes
 the left-moving space-time fermion  number.
Moreover,  $\sigma$ is a holomorphic, isometric
 involution. The first choice leads to orientifold $O9$- and $O5$-planes
whereas the second choice to $O7$- and $O3$-planes.
The generated R-R tadpoles need to be cancelled by the introduction
of the corresponding D-branes. For latter case, the one of primary
interest here,  these are in general
$D7$-branes carrying addition gauge flux and $D3$-branes.
%The massless bosonic spectrum of the ten-dimensional type IIB theory includes
%the dilaton $\phi$ with an axion $C_0$, the metric $g$, an NS-NS two-form
%$B_2$, RR forms $C_2$ and $C_4$, which has a self-dual field strength in the
%R-R sector.
The $(-)^{F_L}\,\Omega_p\, \sigma$  invariant states in four-dimensions
are listed in table \ref{tableprojection}.
\begin{table}[H]
  \centering
  \begin{tabular}{c|c|c|c}
     & $(-)^{F_L}$  &  $\Omega_p$   & $\sigma^*$ \\
    \hline \hline
    $\phi$ & $+$ & $+$ & $+$ \\
    $C_0$ & $-$ & $-$ & $+$ \\
    \hline
    $g_{\mu \nu}$ & $+$ & $+$ & $+$ \\
    $B_2$ & $+$ & $-$ & $-$ \\
    \hline
     $C_2$ & $-$ & $+$ & $-$ \\
     $C_4$ & $-$ & $-$ & $+$ \\
  \end{tabular}
  \caption{\small Orientifold invariant states.}
  \label{tableprojection}
\end{table}
%In the four-dimension effective compactified theory, these ten-dimensional
%fields are expanded out in terms of harmonic forms and only the invariant
%states of the full orientifold projection survive.

Therefore,  the massless states are in one-to-one correspondence
with harmonic forms which are either  even or odd
under the action of $\sigma$.
These do  generate the equivariant  cohomology groups $H^{p,q}_\pm (X)$.
Therefore, the K\"{a}hler form $J$, the
two-forms $B_2$,  $C_2$ and the R-R four-form $C_4$ can be expanded
as
\eq{
J &= t^\alpha\, \omega_\alpha\\
C_2&=c^a\, \omega_a \, ,\quad  B_2= b^a\, \omega_a \\
C_4 &= D_2^{\alpha}\wedge \omega_\alpha +
V^{\tilde{\alpha}}\wedge\alpha_{\tilde{\alpha}}+
U_{\tilde{\alpha}}\wedge\beta^{\tilde{\alpha}} + {\rho}_{\alpha} \, \tilde\omega^\alpha
}
where $\omega_\alpha$ and $\omega_a$ denote a
bases for $H^{1,1}_+(X)$ and $H^{1,1}_-(X)$, respectively.
Similarly, ${\tilde{\omega^\alpha}}$ and  ${\tilde{\omega^a}}$ is a basis of $H^{2,2}_+(X)$ and
$H^{2,2}_-(X)$,
while $(\alpha_{\tilde{\alpha}},\beta^{\tilde{\alpha}})$ is a real
symplectic basis of $H^3_+(X)$.

Since $\sigma^*$ reflects the holomorphic three-form,
in the orientifold one keeps
$h^{2,1}_-(X)$ complex structure moduli $z^{\tilde a}$, which
are complex scalars.
Moreover, $b^a, \, c^a$ and $\rho^\alpha$ are also  scalars, while
$V^{\tilde{\alpha}}$ and $U_{\tilde{\alpha}}$ are space-time one forms
and $D_2^{\alpha}$  a space-time two-form.
Due to the self-duality of the R-R four-form, half of
the degrees of freedom of $C_4$ are removed.
Note that the even component of the Kalb-Ramond field
$B_{+} = b^\alpha \, \omega_\alpha$, though  not a continuous modulus,
can take the two  discrete values $b^\alpha\in\{0, 1/2\}$.
The resulting ${\cal N}=1$ supersymmetric
massless bosonic spectrum is summarized in
Table \ref{table_susy}.

\begin{table}[H]
\centering
\begin{tabular}{|l|l|l|}
\hline
 &  & \\[-0.2cm]
  & $h_-^{2,1}$ & $z^{\tilde a}$ \\
chiral multiplets & $h_+^{1,1}$ & $(t^\alpha, \rho_{\alpha})$ \\
  & $h_-^{1,1}$ & $(b^a, c^a)$\\
   & 1 & $(\phi, C_0)$\\
    &  & \\[-0.2cm]
    \hline
    &  & \\[-0.2cm]
vector multiplet & $h_+^{2,1}$ & $V^{\tilde\alpha}$ \\
&  & \\[-0.2cm]
\hline
 &  & \\[-0.2cm]
gravity multiplet & 1 & $g_{\mu\nu}$ \\[0,2cm]
\hline
\end{tabular}
\caption{${\cal N}=1$ massless bosonic spectrum of Type IIB Calabi Yau
  orientifold
\label{table_susy}}
\end{table}
\noindent
By performing the detailed dimensional reduction from ten to
four dimensions \cite{Lust:2006zg}, one realizes that
the complex bosons in the chiral superfields are given by
the combinations
\eq{
\label{eq:N=1_coords}
S&=i \, C_0+ e^{-\phi} \, , \\
{G}^a &= i \,  c^a - S \, {b}^a \, ,\\
T_\alpha&=\frac{1}{2} \kappa_{\alpha\beta\gamma}\, t^\beta t^\gamma +  \,
  i \left(\rho_\alpha -\frac{1}{2}\kappa_{\alpha a b} \, {c^a b^b}\right)
-\frac{1}{4}\, e^\phi \,  \kappa_{\alpha ab}\, {\bar G}^a {(G+\bar G)}^b\, .
}
The low energy effective action at second order in derivatives is
given  by a supergravity theory, whose dynamics is encoded
in a K\"{a}hler potential $K$, a holomorphic superpotential $W$ and the
holomorphic gauge kinetic functions.
In our case, the K\"{a}hler potential can be expanded as
\eq{
\label{eq:K}
K = - \ln\left(S+{\bar S}\right)
-\ln\left(-i\int_{X}\Omega_3\wedge{\bar\Omega_3}\right)-2\ln\left( {\cal Y}\,
(S, G^a, T_\alpha,...)\right)
}
where ${\cal Y}= \frac{1}{6}{\cal K}_{ABC}\, t^A t^B t^C$ is the volume
of the Calabi-Yau manifold expressed in terms of two-cycle
volumes $t^A$.
The dots in \eqref{eq:K}  denote the potential appearance of other moduli
%with correspoonding shift in dilaton fileds, e.g.
like D3/D7-brane fluctuations (and
hence complex structure moduli which get coupled after including
brane-fluctuations) or Wilson line moduli.
Unfortunately, ${\cal Y}$ is only implicitly given in terms
of the chiral superfields.  It is in general non-trivial
to invert the last relation in \eqref{eq:N=1_coords}.

As we will review in more detail below,
the general schematic form of the superpotential $W$ is given as
\bea
\label{eq:W}
W & &= \int_{X}G_3\wedge\Omega + \sum_{E} {\cal A}_{E}(z^{\tilde a}, G^a, \,{\cal F}_E\,, ...)
\, e^{-\, \pi a_E^\alpha T_{\alpha}}  \nonumber\\
& &= W_0 + W_{np}
\eea
where the first term is the Gukov-Vafa-Witten (GVW)
three-form flux induced tree-level superpotential \cite{Gukov:1999ya} (See \cite{Dasgupta:1999ss,Taylor:1999ii} also for related work).
The second term denotes a sum over  non-pertur\-ba\-tive corrections coming from
Euclidean $D3$-brane instantons or gaugino condensation on D7-branes
\cite{Witten:1996bn}. Here, the prefactor does not only contain the one-loop
Pfaffian for fluctuations around the instanton background but also
contributions from so-called (gauge-)fluxed instantons
and Euclidean $D1$-brane instantons.
Again the dots indicate a further dependence on e.g.
D3/D7-brane fluctuations or Wilson line moduli.
From the  K\"{a}hler- and the superpotential one can compute the
${\cal N}=1$ scalar potential
\eq{
V = e^{K}\Biggl(\sum_{I,\,J}{K}^{I\bar{J}} {\cal D}_I W
{\bar{\cal D}}_{\bar J } {\bar W} - 3 |W|^2 \Biggr)
}
where the sum  runs over all moduli.
%For studying the K\"{a}hler moduli dynamics, we will be assuming that complex structure moduli and
%axion-dilaton has been already stabilized supersymmetrically  as ${\cal D}_{c.s.} W = 0, \, {\cal D}_{S} W = 0$.

\subsubsection*{Fluxed $D3$-brane instantons}

Let us provide some more relevant information about the
fluxed $D3$-brane instanton contributions to the superpotential.
Here we essentially follow \cite{Grimm:2011dj, Kerstan:2012cy}\footnote{For more on instanton-corrections to the superpotential, see \cite{Blumenhagen:2009qh}. For recent progress towards the possibility of new-instanton corrections; see \cite{Bianchi:2011qh} for rigidifying the deformation zero modes of a divisor wrapped by E3-instanton, and \cite{Berglund:2012gr} for avoiding the strong constraints from Freed-Witten anomaly \cite{Freed:1999vc} relevant for a non-spin divisor wrapping an E3-instanton.}.

%Recently, in \cite{Blumenhagen:2007sm}, a tension between moduli stabilization and chirality
%has been raised demanding that one has to be more careful about this while making attempts to
%construct realistic models. The basic idea behind this issue is that with the inclusion of a chiral
%observable sector in the compactification scheme, the charged instanton zero modes localizing on the
%intersecting D7-branes with world-volume flux are not allowed to contribute in the Superpotential and
%hence (K\"{a}hler) moduli stabilization which usually based on this contribution to the Superpotential
%gets crucially affected. A possible resolution to this problem has been proposed in \cite{Grimm:2011dj}
%with the inclusion of instanton flux. The authors have argued that instanton flux on Euclidean D3-branes
%can remove these extra zero modes and help in
%alleviating the general tension between moduli stabilization and chirality.

For a single Euclidean $D3$-brane instanton to contribute to the superpotential
it needs to carry  the right zero-mode structure. In particular, the instanton
has to be a so-called $O(1)$ instanton, which means that
is has to  wrap   an orientifold invariant four-cycle,
i.e. $\sigma(D_E)=D_E$. In addition one has the freedom to turn
on a gauge flux  $\tilde{{\cal F}}_E = 2 \pi  \alpha^\prime \, {\cal F}_E
-\iota^* B$ on the brane, where $\iota:D_E\to X$ denotes the inclusion
map of the four-cycle into the Calabi-Yau threefold.
Since the gauge flux is anti-invariant under the  world-sheet
parity transformation $\Omega_p$, the instanton remains to be
$O(1)$ only if the  gauge flux is supported on a  $\sigma$-odd
two-cycle. Therefore, the gauge flux is supported on two-cycles
in $H^{1,1}_-(D_E)$ which can be expanded as
\eq{
 2\pi\alpha^{'} {\cal F}_E = {\cal F}^{a}_E \; \iota^* \omega_{a} +
{\cal F}^v_E
}
where $\iota^* \omega_{a}$ denotes a $\sigma$-odd basis of harmonic two-forms
lying in the image of the pullback $\iota^*$.
The second component ${\cal F}^v_E$ is given by fluxes supported
on two-cycles inside the divisor $D_E$, which are trivial
in the bulk, i.e. they  are in the co-kernel of $\iota^*$.

Such a family of  fluxed $D3$-brane instantons, all
wrapping the divisor $D_E$, contributes to the superpotential
as
\eq{
W_{np}\sim \sum_{{\cal   F}_E} e^{-S_E}\, .
%\qquad  S_E = (2 \pi \alpha^\prime)^{-2} f_E\; .
}
%where $f_E$ denotes the gauge kinetic function of a ficticious
%$D7$-brane wrapping the same four-cycles as the instanton.
 Dimensionally reducing the corresponding DBI and CS actions, one finds
\eq{
 S_E = \pi\, \Big(a_E^\alpha (T_\alpha +  \Delta_{E\alpha})  +  \Delta_E^v\Big)
}
with
\eq{
\Delta_{E\alpha} &= \kappa_{\alpha b c}\, G^b\, {\cal F}_E^c + \frac{S}{2}
\Big(\kappa_{\alpha b c}\, {\cal F}_E^b\, {\cal F}_{E}^c \Big) \, ,\\
 \Delta_{E}^v &= S \int_{D_E} {\cal F}_E^v \wedge {\cal F}_E^v
}
where  $a_E^\alpha = \int_{D_E} \tilde{w}^\alpha$.
%In the above, similar to the case of partition function of M5-brane instantons
%in M/F-theory \cite{Witten:1996bn,Ganor:1996pe}, the summation over
%instanton-flux above results in theta-functions in the instanton partition
%function which is given in terms of VEVs of odd-moduli $G^a$ and
%instanton-flux.
Collecting the various terms, the superpotential can be
written as
\eq{
\label{eq:Wfluxed-inst}
W_{np} = \sum_{{\cal F}_E} {\cal A}_{E}({\cal F}_E) \, e^{-\pi\, a_E^\alpha\, T_{\alpha}
-  \tilde{q}_{Ea}\, G^a}
}
where
\eq{
{\cal A}_{E} ({\cal F}_E) &= {A} \, \exp\left( - \, {\textstyle\frac{S}{2}}
\Big[\pi\, a_E^\alpha \, \kappa_{\alpha b c}\, {\cal F}_E^b \, {\cal F}_E^c  +2\pi
\int_{D_E}  {{\cal F}}_{E}^v \wedge {{\cal F}}_{E}^v\Big]\right) \\[0.2cm]
{\tilde q}_{Ea} &=  \pi\, \kappa_{\alpha a b}\, a_E^\alpha\,  {\cal F}_E^b  \, .
}
Here $A$ denotes the  one-loop determinant for fluctuations
around the  instanton, which only depends  on the complex structure moduli
and the $D7$-brane deformations.
Thus, it can be assumed to be a constant for our analysis.

Following the discussion in \cite{Kerstan:2012cy},  one now introduces
a basis of two-forms of $H^{1,1}_-(D_E)$ satisfying
$\int_{D_E} \omega_M\wedge \omega_N =2\, \delta_{MN}$ with the
index $M=\{m,\hat m\}$.
The extra factor of two is due to the orientifold
projection (see \cite{Kerstan:2012cy}). The two-cycles $\omega_m$
% runs over both the set of pull-back cycles and its
%complement. The extra factor of two is due to the orientifold
%projection (see \cite{}).
are related to the pull-back
two-cycles in $\iota^* H^{1,1}_-(X)$ as $\omega_a=M_a^m\, \omega_m$
where the matrix $M_a^m$ satisfies
\eq{
a_E^\alpha\, \kappa_{\alpha bc}=\int_{D_E} \omega_b\wedge \omega_c
 =2\, M_a^m\, M_b^n\, \delta_{mn}\, .
}
The $\omega_{\hat m}$ denote a basis of the orthogonal complement of
$\iota^* H^{1,1}_-(X)$ in $H^{1,1}_-(D_E)$.

Now, one can expand the instanton fluxes as
${\cal F}_E=\sum_M  f^M\, \omega_M$ with $f^M\in \mathbb Z$
so that the entire instanton generated superpotential can be written
as
\eq{
\label{eq:Wnpfinal}
%W &= \int_{X} G_3\wedge\Omega +
W_{np}&=A \sum_{E} \, e^{-\pi a_E^\alpha T_{\alpha}}
\sum_{f^M \in {\mathbb Z}} \, \exp\Big(-\pi S f^M f^N\delta_{MN}-
    2\pi\, G^m\,f^n \delta_{mn}\Big) \,
}
Due to the diagonal form in the exponential part of eq.(\ref{eq:Wnpfinal}),
it can be further simplified as:
\eq{
\label{oddWnp}
W_{np}&= A_s \sum_{E}e^{-\, \pi \,a^{\alpha}_E  T_{\alpha}} \sum_{f^{m} \in {\mathbb Z}} \,
e^{-\pi \,S \,  {f^m}^2- 2\, \pi \, f^{m} G^m }
}
where  $A_s= A  \sum_{f^{\hat m} \in {\mathbb Z}} e^{- \, \pi\, S\, {f^{\hat m}}^2 }$ is just a ${\cal O}(1)$
constant.
This is the form of the superpotential to be heavily utilized in the
subsequent analysis of axion moduli stabilization. There we will assume
that the zero mode structure  of such an instanton is just right to guarantee
a contribution to $W$.

In general, for stabilizing the odd moduli in a realistic setup which concretely supports a MSSM-like visible sector, one should also examine for the possible D-term potential
coming from the D7-brane fluxes ${\cal F}_A$ turned-on on a stack of D7-branes along the
holomophic divisor $D_A$ and its orientifold image $D_{A'}$. Such a D-term reads as
\bea
\label{D-term}
& & {\cal D}_A =\frac{l_s^2}{2 \pi {\cal V}} \int_{D_A} J \wedge (\iota^* B_2- 2 \pi \alpha' {\cal F}_A) \nonumber\\
& & \hskip2cm = \frac{l_s^2}{4 \pi {\cal V}} t^{\alpha} \left(\kappa'_{\alpha b c}(b^b-{\cal F}_A^b) C_A^c -\kappa_{\alpha\beta\gamma}{\tilde{\cal F}}_A^\beta \, C^\gamma_A\right) 
\eea
where $C_A^{\alpha, a}=N_A\int_{D_A^\pm} {\tilde \omega}^{\alpha,a}, \, (D_A^\pm=D_A \cup (\pm D_{A'}))$ are
the wrapping number along the basis  of
$H_4^\pm(X)$ and $\kappa_{ABC} := \int_{CY_3} \omega_{A} \wedge \omega_B \wedge \omega_C$ with $A=\{\alpha, a\}$ gives the intersection numbers for even/odd sectors. 
In general, several $\kappa_{\alpha b c}$ intersection numbers can be non-zero, then $b^a$ moduli are stabilized at tree level by
requiring  D-flatness condition\footnote{For example, one way to impose the $D$-flatness conditions is to set two-cycle volumes $t^{\alpha}$ appearing in \eqref{D-term} to zero. This leads to models of D-branes at singularity in \cite{Cicoli:2012vw} where only self-intersecting (shrinkable) del-Pezzo divisors have been exchanged under involution $\sigma$. }. However, this may not be always the case; for example, to generate the FI-term as given in \eqref{D-term}, one requires a
 $U(1)$ group on the D7-brane configuration which may not be necessarily met. %Such a situation arises when we place eight $D7$-branes on top of the $O7$-plane, and subsequently get an $SO(8)$ group structure on these $D7$-branes.
In case of a $U(1)$ gauge group being present with certain brane configuration, it can also happen that the two-cycle $\omega_b$ intersects with the $\omega_c$ dual to $D_A^-$ divisor trivially, i.e $\kappa'_{\alpha b c} = 0$. It is worth to mention that $\kappa'_{\alpha b c}=0$ does not mean that the intersection number between E3-instanton divisor and odd cycle has to vanish ($\kappa_{\alpha b c} \neq 0$) unless the $D7$-brane appearing in (\ref{D-term}) wraps the same divisor. {\it Here, we stress that our main motivation is to investigate F-term moduli stabilization with the inclusion of instanton flux effects in a toy model, and therefore in the present work, we assume that concrete setups with desired divisor intersections and allowed fluxes can be constructed. Further, as our investigations are based on a phenomenologically oriented approach, we do not intend to explicitly address the issue of supporting a concrete MSSM-like visible sector. 
%As a by-product of our moduli stablization investigations, we find that although $c^a$ axions are indeed lightest in weak coupling limit, it is not a possible candidate for a QCD axion within the simple moduli stablization standard approach we follow.
} %However, since in this paper we do not intend to realize a realistic model by including visible sector explicitly, 
Therefore, we can choose the configuration in which $D7$-brane does not wrap the instanton divisor and results in  ${\tilde q_{E_a}} \neq 0$ while $\kappa'_{\alpha b c}=0$. %Then we can ask for the possibility to stabilize odd moduli only by F-term. With such possibilities, $b^a$ moduli are not stabilized by $D$-term. 

\section{Extended LARGE Volume Scenario}
\label{sec_Even-axions}
In this section, we discuss the moduli stabilization in an extended LARGE volume setup with
the inclusion of involutively odd axions in the context of type IIB orientifold compactification.
%We will also estimate the axion decay constants and masses of various even/odd axions. %  in order to
%investigate the possibility of realizing QCD axion candidate.
Let us start with briefly reviewing
the standard features of the minimal LARGE volume setup. We assume that all the complex structure
moduli as well as axion-dilaton are supersymmetrically
stabilized at the perturbative stage by background-flux superpotential $W_0$ via  ${\cal D}_{\rm c.s.} \,
W _0\, = 0 = \, {\cal D}_{S} W_0$.
This remains justified with the inclusion of non-perturbative contributions as long as
the overall volume of the Calabi-Yau space remains sufficiently large. For stabilizing the
K\"ahler moduli, one starts with the following form of  K\"ahler potential and superpotential,
\eq{
\label{eq:KW+simplest+LVS}
 K &= - 2 \, \ln {\cal Y} \,, \, \, \,\,\,  W = W_0 + \sum_{s =2}^{h^{1,1}_+} \, A_s \, e^{-a_s T_s}\,,
}
where ${\cal Y}= {\cal V}(T_\alpha)+C_{\alpha^\prime}$ such that
\bea
{\cal Y}
= \xi_b (T_b+\bar{T}_b)^{\frac{3}{2}}-\sum_{s =2}^{h^{1,1}_+} \, \xi_s
(T_s+\bar{T}_s)^{\frac{3}{2}} +
C_{\alpha^\prime}\, . \nonumber
\eea
Here, we consider the ansatz of multi-hole swiss-cheese structure in the Calabi-Yau volume with a shift $C_{\alpha^\prime}$ which denotes  the perturbative
${\alpha^\prime}^3$-correction  given as
$
C_{\alpha^\prime} = - \frac{\chi({\cal M}) \,
  ({{\tau}-\bar\tau})^{\frac{3}{2}} \zeta(3) }{4 (2 \pi)^3 \,
  ({2i})^{\frac{3}{2}}}
$.
This ${\alpha^\prime}^3$-correction breaks the no-scale structure \footnote{In the meantime, there have been proposals for string-loop corrections \cite{Berg:2007wt,Cicoli:2008va} as well as `new' $\alpha^\prime$-corrections \cite{Anguelova:2010ed, Grimm:2013gma, Pedro:2013qga} (see also \cite{GarciaEtxebarria:2012zm} for a related progress in ${\cal N}=2$ F-theory compactifications). However, an `extended' no-scale structure has been observed making the LARGE volume scenarios more robust. From a field theoretic approach, similar structure has been observed earlier for certain form of corrections to the K\"ahler potential \cite{vonGersdorff:2005bf}.}, and the ansatz \eqref{eq:KW+simplest+LVS} results in the following form of $F$-term effective scalar potential
\bea
\label{eq:VLVSsimpest}
& & \hskip-2cm V^{\rm LVS}({\cal V};\{\tau_s,\rho_s\}) =\frac{3 {\, {\cal C}_{\alpha^\prime}} \,
|W_0|^2}{2 \, {\cal V}^{3}}+ \sum_{s =2}^{h^{1,1}_+} \,
\frac{2 {\sqrt 2}\,a_s^2 \,A_s^2\, e^{-2 {a_s \tau_s}} \sqrt{{\tau_s}}}{3 \, \xi_s \, {\cal V}}\nonumber\\
& & \hskip5cm + \sum_{s =2}^{h^{1,1}_+} \, \frac{{4 \,a_s} {\,{A_s}}\, e^{-{a_s \tau_s}}\,
{\tau_s} \cos [a_s \rho_s] {W_0}}{{\cal V}^2}.
\eea
This potential stabilizes the overall volume mode at exponentially large value in terms
of stabilized value of the `small' divisor volume $\ov{\cal V}\sim |W_0| \exp(a_s\, \ov\tau_s)$
where $\ov\tau_s \sim (C_{\alpha^\prime})^{2/3}$. One realizes a non-supersymmetric $AdS$ minimum
which can be uplifted to a de-Sitter minimum via various uplifting mechanisms \cite{Kachru:2003aw,Balasubramanian:2004uy,Westphal:2006tn,Burgess:2003ic,Saltman:2004sn,Cicoli:2012fh}. Further,
the leading order contributions to the decay constants for all
the axions can be estimated in large volume limit to be,
\begin{equation}
 f_{\rho_b} = \frac{\sqrt6 \, \xi_b^{2/3}}{{\cal V}^{2/3}} \sim {\cal V}^{-2/3}, \, \, {\rm and } \, \, \,
f_{\rho_s} =
\frac{{\sqrt{3 \, \xi_s}}}{(2 \tau_s)^{1/4} {\cal V}^{1/2}} \sim {\cal V}^{-1/2} \, \, \, \, \, \,
\forall s \in \{2, ..., h^{11}_+\}
\end{equation}
After looking at the eigenvalues of the mass-squared matrix $M_{ij} \equiv \sum_k \frac{1}{2} (K^{-1})_{ik} V_{kj}$,
one gets the following leading order contributions for moduli masses (evaluated at the minimum),
\bea
& & M_{\cal V} \sim \frac{\delta}{{\cal V}^{3/2}}, \, M_{\tau_s} \sim \frac{\delta}{{\cal V}}, \, M_{\rho_b} = 0,\,\,\,\ M_{\rho_s} \sim \frac{\delta}{{\cal V}} \,; \, \,  \, \forall s \in \{2, ..., h^{11}_+\}.
\eea
where $\delta \sim \sqrt{\frac{g_s \, e^{K_{CS}}\, |W_0|^2}{8\pi} }$. It is important
to note that the axionic direction $\rho_b$ corresponding to the non-local (so-called `big') divisor remains
flat. Furthermore, one can lower the decay constant $f_{\rho_b}$ naturally
(in large volume limit) to get the correct order of magnitude for QCD axion, and so the $\rho_b$ axion has
appeared to be quite attractive for this purpose \cite{Conlon:2006tq}\footnote{See \cite{Cicoli:2012sz, Higaki:2011me, Marsh:2012nm} also for recent progress with more phenomenological approach.}. %However, one should
%make sure that this axion should not be eaten up via St\"uckelberg mechanism.
%In a more rigorous study \cite{Cicoli:2012sz},  the decay constant have been investigated
%for (non-)local $C_4$ axions type, however, the origin of the non-zero small masses relevant
%for a QCD axion has not been addressed. In the present article, our main focus is to
%investigate $C_2$ axion for being a QCD axion candidate via looking at the possibility
%of generating small masses through fluxed-instanton superpotential contributions.

Now, let us consider an extension of the simplest LARGE volume scenario with
the inclusion of a single involutively odd modulus $G^1$. 
%In the context of type IIB orientifold
%framework, a holomorphic involution $\sigma$, which could result in non-trivial odd sector for
%(1,1)-cohomology, i.e. $h^{1,1}_-(CY_3/\sigma) \ne0$, is needed. This can be obtained by exchanging
%two ``nontrivial identical'' four-cycles,which are divisors with different
%GLSM (Gauged Linear Sigma-Model) charges
%  intersecting with the Calabi-Yau hypersurface to the same surface and hence with the same Hodge number
%\footnote{A systematic analysis exemplifying concrete setups with odd moduli in type IIB orientifold
%framework is under progress \cite{Gao:2013} in which all the Calabi Yau threefolds with $h^{11}(CY_3)\le4$
%with  possible exchange involutions, $\sigma: x_i \leftrightarrow x_j$, have been classified.}.
In order to support the odd modulus, a non-zero component in (1,1)-cohomology class on the Calabi-Yau threefold under some holomophic involution $\sigma$ is needed, i.e $h^{1,1}_-(CY_3/\sigma) \neq 0$.
In \cite{Gao:2013pra},  we scanned through the toric Calabi-Yau threefolds with $h^{1,1}(CY_3) \leq 4$ and studied two kinds of involutions, namely divisor exchange involution and divisor reflection,  which can result a non-trivial odd (1,1)-cohomology. 
In the presence of a single odd modulus $G^1$, the superpotential (\ref{oddWnp}) including the non-perturbaive
fluxed-instanton contribution becomes
\bea
\label{oddW0}
W = \int_{X} G_3 \wedge \Omega + A_s e^{-\, a_s  T_{s}} \sum_{f^{1} \in {\mathbb Z}} \,
e^{-\pi \,S \,  {f^1}^2- 2\, \pi \, f^{1} G^1 }
\eea
where $A_s= A  \sum_{f^{\hat m} \in {\mathbb Z}} e^{- \, \pi\, S\, {f^{\hat m}}^2 }$.
Again, we assume that all the complex structure moduli and axion-dilaton are stabilized
by Gukov-Vafa-Witten superpotential perturbatively. For simplicity, we also assume that the
background flux are tuned such that  RR scalar is set to zero, ${\bar C_0}=0$. Subsequently,
the non-perturbative term in the superpotential (\ref{oddW0}) can be written in terms of simplified
elliptic theta function $\theta(G^1)$. The appearance of theta-function as a holomorphic pre-factor of a standard instanton correction (to the superpotential) has been also argued in \cite{Grimm:2007xm}. The same was based on modular completion arguments assuming that a subgroup of $SL(2,{\mathbb Z})$ survives after orientifold truncation. 

For a generalized LARGE volume setup, we proceed with the following ansatz for the K\"ahler
potential and the superpotential
\bea
\label{oddW}
& &K= -2\, {\rm ln} \,{\cal Y} = -2 \, {\rm ln} \, \left(\xi_b \, \Sigma_b^{3/2} -\xi_s \, \Sigma_s^{3/2} + {{\cal C}_{\alpha^\prime}}\right), \\
& &W=  W_0 + A_s \sqrt{ g_s} \,\, e^{-\, a_s T_s} \left(e^{\,g_s \, \pi\, G^1 G^1} \, \, {\cal \theta}_3\left[ g_s \pi G^1, e^{-g_s \pi} \right] \right). \nonumber
\eea
where
\begin{eqnarray}
\label{eq:Yodd1}
 & & \Sigma_b = T_b + {\bar T}_b + \frac{ \kappa_{b11}}{2(S+ {\bar S})}({G}^1+{\bar G}^1)({G}^1+{\bar G}^1)\\
 & & \Sigma_s = T_s + {\bar T}_s + \frac{ \kappa_{s11}}{2(S+ {\bar S})}({G}^1+{\bar G}^1)({G}^1+{\bar G}^1)\nonumber
\end{eqnarray}
Depending on the  possible intersections of various even/odd four cycles,
we consider  two cases for stabilizing {\it all} the even/odd moduli using $F$-term contributions.
These two cases are also of interest because of the different volume scaling in the leading order
axion decay constants as we will see later.

\subsection*{Case-I : $\kappa_{b11} \neq 0$}
\begin{figure}[H]
\centering
\includegraphics[scale=0.4]{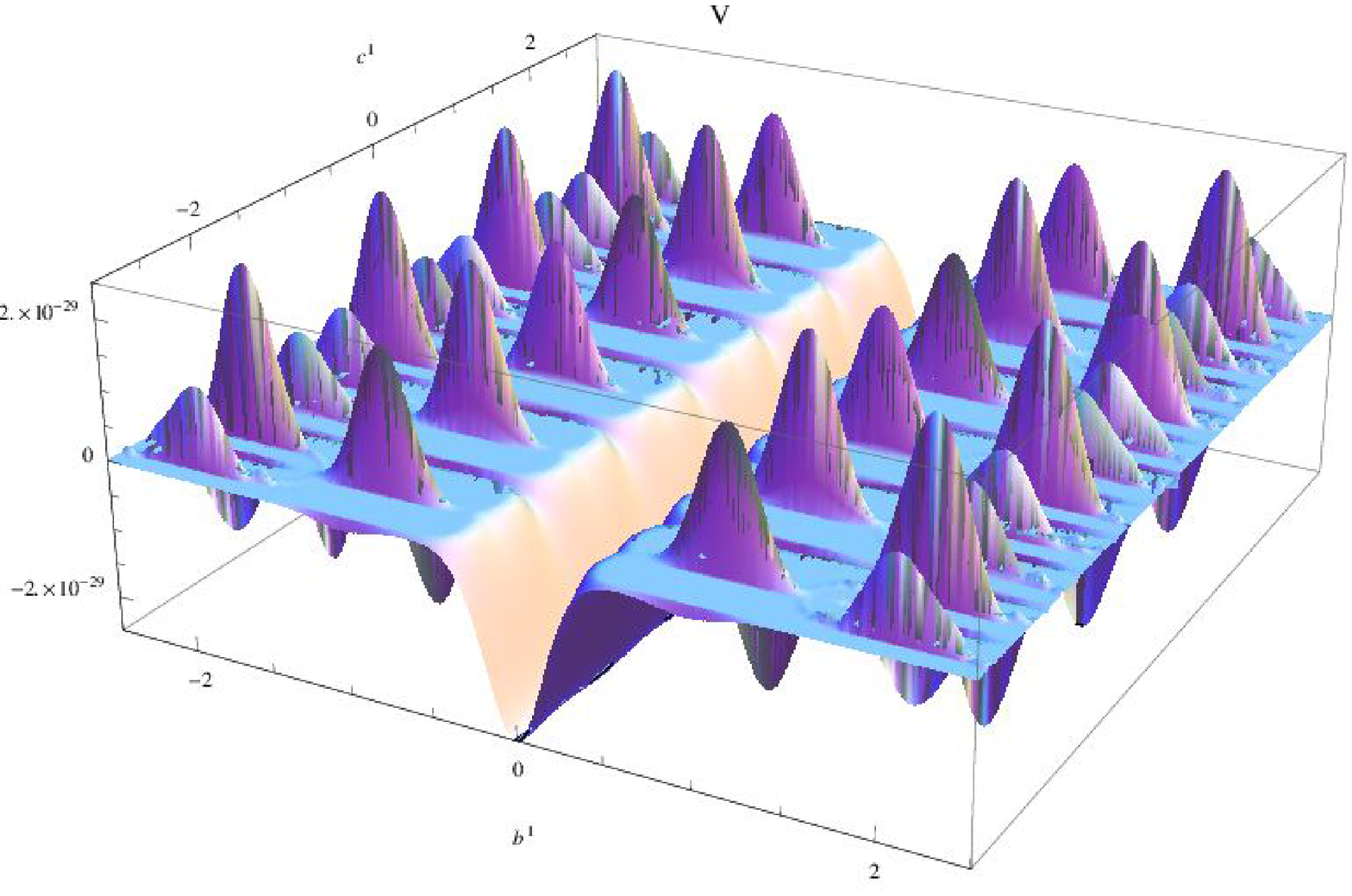} \,\,\,\,\,
\includegraphics[scale=0.4]{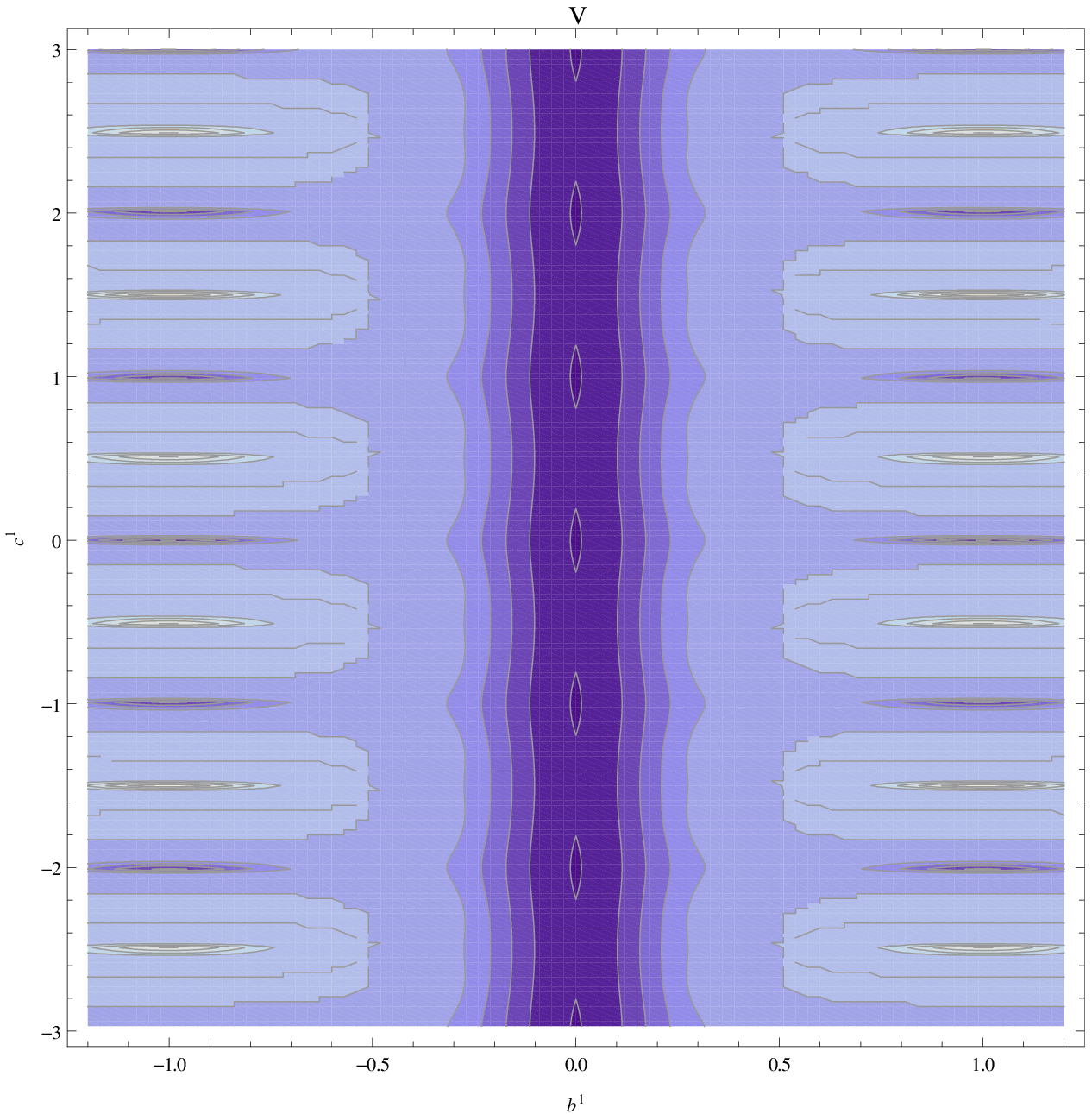}
\caption{ The quasi-periodicity  of the scalar potential in the odd-axion direction $b^1$ and $c^1$
after stabilizing all the other even moduli in a consistent way.}
\label{oddperiodic}
\end{figure}

\begin{figure}[H]
\centering
\includegraphics[scale=0.5]{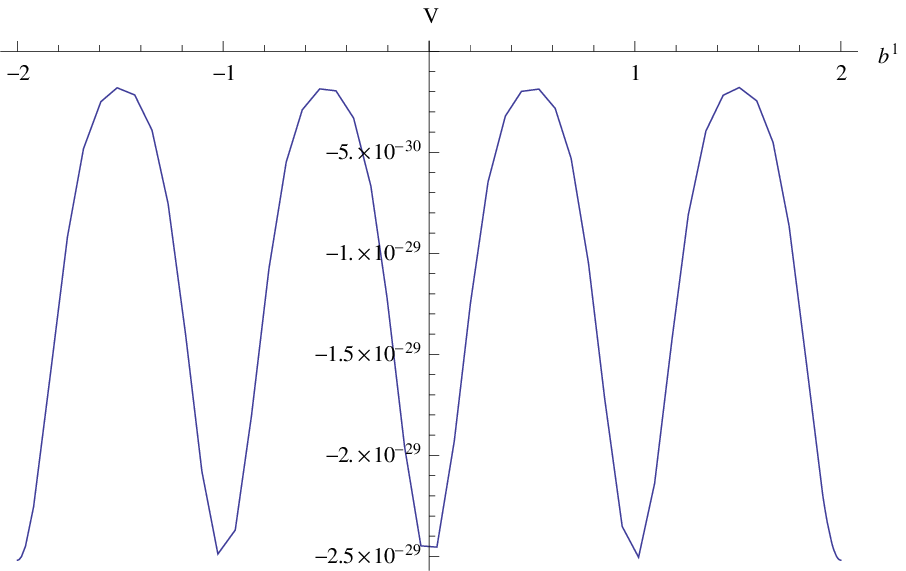} \,\,\,\,\,
\includegraphics[scale=0.5]{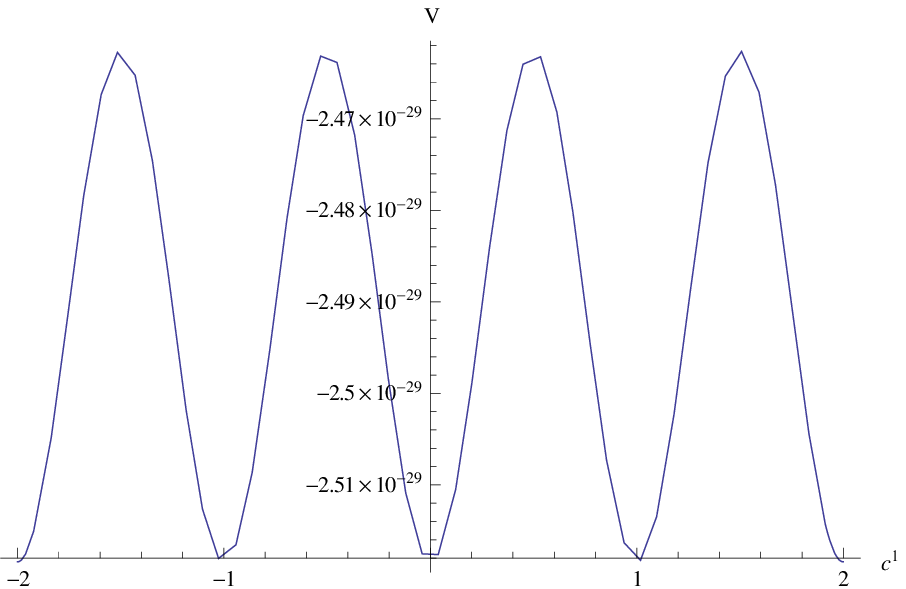}
\caption{ The periodicity of the scalar potential in $b^1$ direction
for $c^1=0$, and in $c^1$ direction for $b^1=0$ after stabilizing all the other even moduli.}
\label{oddperiodic2}
\end{figure}

Let us assume that the so-called big divisor has non-zero intersections with the odd cycles,
i.e. $\kappa_{b11} \ne0$. Utilizing the ansatz (\ref{oddW})-(\ref{eq:Yodd1}) for the K\"ahler potential
and superpotential, the leading order contributions to
the F-term scalar potential can be collected
in three types of terms as under,
%(simialr to the standard large volume potential)
\bea
\label{oddV1}
& & \hskip-0.5cm V({\cal V}, \tau_s; \rho_s, b^1, c^1) = V_{\alpha'} + V_{np1} + V_{np2} \, \, \, \, \, \, {\rm where }\nonumber\\
& & \nonumber\\
& & \hskip-0.5cm V_{\alpha^\prime} = \frac{3 \, {{\cal C}_{\alpha^\prime}} \,  |W_0|^2}{2\,  {\cal V}^3} \\
& & \hskip-0.5cm V_{np1}=\frac{2 a_s A_s \tau_s \sqrt{g_s} W_0}{{\cal V}^2} \times {\rm exp}\left[\frac{a_s}{2 g_s}(\kappa_{s11}b^1b^1
-2 g_s \tau_s )+\frac{ \pi}{g_s} (b^1-i g_s c^1)^2 \right] \nonumber\\
& & \hskip-0.5cm \hskip1cm \times \biggl\{ {\rm cos}(a_s \rho_s) \left(\bar\Theta(b^1,c^1) e^{i 4 \pi a_s b^1 c^1} +\Theta(b^1,c^1)
\right) \nonumber\\
& & \hskip-0.5cm \hskip2cm  - i \, {\rm sin}(a_s \rho_s) \left(\bar\Theta(b^1,c^1) e^{i 4 \pi a_s b^1 c^1}
-\Theta(b^1,c^1) \right) \biggr\} \nonumber
\eea
\bea
& & \hskip-0.5cm V_{np2}=\frac{2\sqrt{2 \tau_s} g_s a_s^2 A_s^2  }{3 \,\xi_s} |\Theta(b^1,c^1)|^2
\times {\rm exp}\left[ \frac{a_s  (\kappa_{s11} b^1 b^1-2 g_s \tau_s)}{g_s}
+ \frac{2  \pi}{g_s}({c^1}^2 g_s^2- {b^1}^2) \right]\nonumber
%& & V_{np1} = \frac{2 {a_s} {\tau_s} {W_0}}{{\cal V}^2} \times \exp\left[\frac{1}{4} \left(\frac{2 {a_s}}{g_s}
%\left(\kappa_{s11}b^1 b^1-2 g_s {\tau_s}\right)+\frac{k_{s1}^2 ({b^1}-i {c^1} {g_s})^2}{\zeta_s  {g_s}}\right)\right]\nonumber\\
%& & \hskip 1in \times \biggl\{ i \sin ({a_s} {\rho_s}) \left({\bar\Theta}(b^1,c^1)-\Theta(b^1,c^1)  e^{\frac{i {b^1} {c^1} k_{s1}^2}{\zeta_s}}\right)\nonumber\\
%& & \hskip1.5 in +\cos ({a_s} {\rho_s}) \left({\bar\Theta}(b^1,c^1)+\Theta(b^1,c^1) \, e^{\frac{i {b^1} {c^1} k_{s1}^2}{\zeta_s }}\right)\biggr\} \nonumber\\
%& & V_{np2} = \frac{2 \, \sqrt{2}  \, a_s^2 \, \sqrt{{\tau_s}}}{3 \, \xi_s \,{\cal V}}\left|\Theta(b^1,c^1)\right|^2 \nonumber\\
%& & \hskip1.5in \times \exp{\left[\frac{k_{s1}^2 ({b^1}^2-{c^1}^2 g_s^2)}{2 \zeta_s {g_s}}+\frac{{a_s}}{g_s} \left({\kappa_{s11} b^1 b^1}-2 g_s {\tau_s}\right)\right]} \nonumber
\eea
where $\Theta(b^1,c^1)=\theta_3\left[ -b^1 \pi +i \,c^1 g_s \pi, e^{-g_s \pi} \right]$.
There are several extrema in the axionic directions due to periodicities appearing in the potential \eqref{oddV1}, and the generic extremization conditions are quite coupled.
However, after utilizing one extremizing condition into another, the most simplest local extremum  of
the scalar potential \eqref{oddV1} can be collectively described  by intersection of the following hypersurfaces
in moduli space. %\footnote{It is good to observe that
%for the extremum values of odd-axions such that  $b^1 = 0 = c^1$, the conditions (${\cal D}_{c.s.}W_0 =0$)
%extremizing the complex structure moduli remains the same.}
\eq{
\label{odd-extreme}
a_s \rho_s &= N \pi, \qquad b^1 = 0, \qquad c^1 = 0, \qquad
{\cal C}_{\alpha^\prime} = \frac{32 \sqrt{2} \, a_s \, \xi_s\, \, \tau_s^{\frac{5}{2}} (-1 + a_s \tau_s)}{(-1 + 4 a_s \tau_s)^2}
\,,\\[0.1cm]
\hskip1cm W_0 &= -\frac{\, a_s A_s \, e^{-a_s \tau_s}\, {\cal V} \, (-1 + 4 a_s \tau_s) \, \sqrt{g_s} \Theta(0)}{6 \, {\sqrt 2} \,
  \xi_s  \, \sqrt{\tau_s}\, (-1 + a_s \tau_s)}. \,
}
where $\Theta(0) = \theta_3[x,e^{- g_s \pi}]_{x=0}$.
From eq.(\ref{odd-extreme}), one finds that similar to the standard LARGE volume
scenario, the $\tau_s$ stabilization condition can get decoupled from ${\cal V}$
dependence and results in $\ov\tau_s
\sim (C_{\alpha^\prime})^{2/3}$.  Subsequently, the overall volume ${\cal V}$ gets stabilized
at an exponential large value via  $\ov{\cal V}\sim \exp(a_s\, \ov\tau_s)$.
The scalar potential at this non-susy $AdS$ minimum (\ref{odd-extreme}) is simply given as,
\bea
\label{eq:AdS}
V_{\rm AdS-min} = -\frac{24 \sqrt{2} \,  \xi_s \, \tau_s^{3/2} {|W_0|}^2 \, ({a_s} \,  {\tau_s}-1)}{{\cal V}^3 \, (1-4 {a_s} {\tau_s})^2}
\eea
It is worth to recall that in the above discussion, we have considered only the simplest minimum for which the extremization
conditions could be analytically solved. In fact, there are many extrema in the odd moduli directions
due to the quasi-periodic property of the inverse elliptic theta function.
An easy way to illustrate such property is to show the section of the scalar potential as a function
of odd-axionic modulus after stabilizing
all the other even moduli in a consistent way \footnote{Form eq.(\ref{oddV1}) one can see that
the volume and $\tau_s$ moduli  couples to $b^1$ and $c^1$ in a complicated way. In
general, it is hard to get a simple expression to show explicitly how  the stabilization conditions for
moduli ${{\cal V}}$ and $\tau_s$ depend on the odd moduli.
However, we can solve these coupled extremization conditions numerically to get a potential
for $b^1$ and $c^1$ moduli.}.
Using the following sampling of model dependent parameters in Table.\ref{odd1-case-1},
the scalar potential eq.(\ref{oddV1})
are shown in Figure \ref{oddperiodic}, \ref{oddperiodic2} where the quasi-periodicity in
both ($b^1$ and $c^1$) directions are observed.
\begin{table}[H]
  \centering
  \begin{tabular}{|c||c|c|c|c|c||c|c|c|c|}
  \hline
   Model & $C_{\alpha^\prime}$  & $\kappa_{b11}$  & $\kappa_{s11}$ & $\xi_b$ & $\xi_s$  & $W_0$  & $a_s$ &$A_s$  & $g_s$     \\
    \hline \hline
   B1  & 1.697 & -1 & -1 & $\frac{1}{9\sqrt2}$ & $\frac{1}{9\sqrt2}$ & -0.1 & $2\pi$ & 1 & 0.5  \\
     \hline
 B2  & 1.697 & -1 & -1 & $\frac{1}{9\sqrt2}$ & $\frac{1}{9\sqrt2}$ & -0.1 & $2\pi$ & 0.5 &  0.1 \\
     \hline
  \end{tabular}
   \caption{Sampling of the model dependent parameters.}
   \label{odd1-case-1}
 \end{table}
%\vskip-1cm
\subsubsection*{Axion decay constant and mass matrix}
\begin{figure}[H]
\centering
\includegraphics[scale=1]{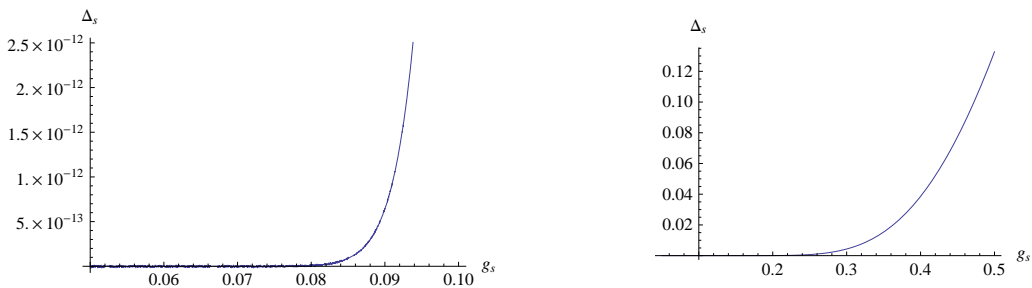}
\caption{The estimate of suppression factor $\Delta_s$ with different values of string coupling $g_s$.}
\label{Delta_s}
\end{figure}

Let us look at the axion decay constant and the masses of various moduli at the non-supersymmetric minimum.
Utilizing the K\"ahler metric, all kinetic terms for the respective moduli can be  written as
\eq{
{\cal L}_{\rm kinetic}({\cal V},\tau_s, \rho_b, \rho_s; b^1, c^1) \equiv
K_{I \bar J} (D_{\mu} T_I) ({\bar D}^\mu {\bar T}_{\bar J}) \,,}
In the basis of real moduli $\{{\cal V},\tau_s, \rho_b, \rho_s; b^1, c^1\}$, the
kinetic matrix (see Appendix \ref{appendixOdd})
is found to be block diagonal in both even and odd sector. The axionic sector of the kinetic matrix
 implies that the leading order contributions to the decay constants for
all the axions can be estimated to be,
\bea
\label{eq:decayOdd}
& & f_{\rho_b} = \frac{\sqrt6 \, \xi_b^{2/3}}{{\cal V}^{2/3}} \sim {\cal V}^{-2/3}, \, \, \, f_{\rho_s} = \frac{{\sqrt{3 \, \xi_s}}}{(2 \tau_s)^{1/4} {\cal V}^{1/2}} \sim {\cal V}^{-1/2} \\
& & f_{b^1} = \frac{\sqrt{-3 \, \kappa_{b11}} \, \xi_1^{1/3}}{\sqrt{2 \,g_s} {\cal V}^{1/3} } \sim {\cal V}^{-1/3}, \, \, \, f_{c^1} = \frac{\sqrt{-3 \,g_s \, \kappa_{b11}} \, \xi_1^{1/3}}{\sqrt2 \,{\cal V}^{1/3} } \sim {\cal V}^{-1/3}. \nonumber
\eea
This shows that the positive definiteness of the kinetic matrix of odd axionic sector demands that $\kappa_{b11} <0$.
Now, let us investigate the squared-mass matrix evaluated at the minimum eq.(\ref{odd-extreme}),
\bea
\label{eq:squaredMassMatrixOdd}
{\cal M}_{ij} =
\left(\begin{array}{lll|ll}
&&&&\\
 \frac{\beta_1}{{\cal V}^{3}}& \hskip 1cm  \frac{\beta_2}{{\cal V}^2} & \hskip 1cm 0 & \hskip 1cm 0 & \hskip 1cm 0 \\
 &&&&\\
 \frac{\beta_4}{{\cal V}^3}& \hskip 1cm \frac{\beta_5}{{\cal V}^{2}} & \hskip 1cm 0 & \hskip 1cm 0 & \hskip 1cm 0 \\
 &&&&\\
 0& \hskip 1cm 0& \hskip 1cm  \frac{\gamma_1}{{\cal V}^{2}} & \hskip 1cm 0 & \hskip 1cm 0 \\
&&&&\\
\hline
 &&&&\\
\hskip .1cm 0 & \hskip 1cm 0 & \hskip 1cm 0 & \hskip 0.5cm M_{b^1 \, b^1}& \hskip 1cm 0\\
&&&&\\
\hskip .1cm 0 & \hskip 1cm 0 & \hskip 1cm 0 & \hskip 1cm 0 & \hskip 0.5cm M_{c^1 \, c^1}
\end{array}
\right)\,.
\eea
The upper left $3\times3$ block corresponds to the even-moduli sector $\{{\cal V}, \tau_s, \rho_s \}$
and reproduces the standard LARGE volume results without odd-axions.  The lower right  $2\times2$
block corresponds to the odd-moduli sector $\{b^1, c^1\}$.  Before we analyze the odd axion mass in detail, let us recall from the superpotential expression (\ref{oddW0}), that in the absence of instanton-flux, the $c^1$ modulus direction is flat as the theta-function appearance in the superpotential disappears,
however, the $b^1$ axionic flatness is lifted even in the absence of instanton-flux because of its implicit appearance in chiral coordinate $T_s$ through the non-perturbative exponential suppression. For the eigenvalues of the mass-squared matrix (\ref{eq:squaredMassMatrixOdd}), one gets the following volume scalings in the moduli masses evaluated at the minimum,
\bea
\label{eq:Mass-odd1}
& & M_{\cal V} \sim \frac{\delta}{{\cal V}^{3/2}}, \,M_{\tau_s} \sim \frac{\delta}{{\cal V}}, \, ; \, \, M_{\rho_b} = 0, \, M_{\rho_s} \sim \frac{\delta}{{\cal V}},\nonumber\\
& & \,  M_{b^1} \sim \frac{\delta \, \Delta_1({\cal F})}{{\cal V}^{\frac{7}{6}}}, \, M_{c^1} \sim \frac{\delta \, \Delta_2({\cal F})}{{\cal V}^{\frac{7}{6}}}
\eea
where $\delta \sim \sqrt{\frac{g_s \, |W_0|^2}{8\pi}}$. The first line represents the expected
results of even-moduli sector and volume scalings are as per expectations \cite{Conlon:2007gk}.  The $\Delta_i({\cal F})$'s appearing in odd axionic masses are introduced in place of multiplicative factors having a theta-function dependence, and are given as,
\bea
\label{delta}
& & \Delta_1({\cal F}) \sim {\cal O}(1) \, \left(g_s \pi^2 \Theta''(0)+ 2 \pi \Theta(0) +a_s \kappa_{s11}\Theta(0)\right)^{1/2}\nonumber\\
& & \Delta_2({\cal F}) \sim {\cal O}(1) \, \left(g_s \pi^2 \Theta''(0)+2 \pi \Theta(0) \right)^{1/2}
\eea
where $\Theta''(0) = \partial_x^2 \, \theta_3(x, e^{-\pi \, g_s})_{|_{x=0}}$. Here, it is worth to mention that $\Delta_i({\cal F})$ does not have explicit flux dependence as fluxes on the instanton divisor are already summed over, even then we mention ${\cal F}$ to keep one reminded that such theta-function contributions are rooted into the instanton flux effects. Further, the naive
volume scalings in mass estimates (\ref{eq:Mass-odd1}) imply that odd-axions are heavier than the overall volume mode. However, the analytic expressions (\ref{delta}) of $\Delta_i({\cal F})$s show that
$|\Delta_1({\cal F})| \sim {\cal O}(1)$ while $|\Delta_2({\cal F})| \ll {\cal O}(1)$ for natural model
dependent parameters. The reason for the same is a crucial multiplicative factor appearing in $\Delta_2({\cal F})$, which is $(2 \theta_3[0, e^{-\pi \, g_s}] + g_s \, \pi \, \theta_3''[0, e^{-\pi \, g_s}])^{1/2}$, as seen from (\ref{delta}).
This is a reasonable amount of suppression of the order $10^{-12}$ for $g_s \sim 0.1$ in mass-squared value of the $c^1$ axion. This happens because of a fine cancellation in two pieces of $\Delta_2$. Let us make it explicit by taking some numerical values,
\begin{table}[H]
  \centering
  \begin{tabular}{|c||cccccc|}
\hline
  $g_s$  & 0.05 & 0.1 & 0.2  &  0.3 & 0.4 & 0.5 \\
    \hline \hline
  $2 \theta_3[0, e^{-\pi \, g_s}]$  & 8.94427 & 6.32456 & 4.47214 & 3.65169 & 3.16473 & 2.83899 \\
\hline
$g_s \, \pi \, \theta_3''[0, e^{-\pi \, g_s}]$  & -8.94427 & -6.32456 & -4.47209 & -3.64736 & -3.12617 &-2.70624 \\
\hline
  \end{tabular}
  \end{table}
The suppression factor in the squared-mass values of $c^1$, which is given as $\Delta_s = (2 \theta_3[0, e^{-\pi \, g_s}] + g_s \, \pi \, \theta_3''[0, e^{-\pi \, g_s}])$), gets more clear from the Figure \ref{Delta_s}.

However, for larger values of string coupling, the factor $\Delta_s$ becomes order one implying
that the mass of $c^1$ axion will be larger than that of overall volume mode. %This makes the
%situation worse from the point of view of our motivation for $c^1$ axion as a QCD axion candidate. 
Thus, for $c^1$ axion to be the lightest, one has to keep string coupling $g_s $
small enough such that $\Delta_s < {\cal O}({\cal V}^{-2/3})$.
 Utilizing the sampling given in
Table \ref{odd1-case-1}, the eigenvalues of mass-squared matrix are shown in Table \ref{odd2case-1}.

\begin{table}[ht]
  \centering
  \begin{tabular}{|c||cc||c|c|cc|}
\hline
  Model  & $\ov{\cal V}$ & $\ov{\tau}_s$ & $m_{\cal V}^2$  &  $m_{\tau_s}^2 \sim m_{\rho_s}^2$ & $m_{b^1}^2$ & $m_{c^1}^2$ \\
    \hline \hline
  B1  & $3\times10^{8}$ & $3.93$ & $3.4 \times 10^{-28}$ & $3.2 \times 10^{-16}$ & $6.9 \times 10^{-22}$ & $3.4\times 10^{-23}$ \\
\hline
B2  & $1\times10^{8}$ & $3.93$ & $3.3 \times 10^{-28}$ & $3.1 \times 10^{-16}$ & $7.2 \times 10^{-22}$ & $1.5\times 10^{-36}$ \\
\hline
  \end{tabular}
  \caption{\small Stabilized values of (divisor) volume moduli along with the eigenvalues of mass-squared matrix
  (evaluated at the minimum) in $Mp = 1$ units. The respective stabilized values for the axions are $\ov\rho_s = 0 = b^1 = c^1$.}
  \label{odd2case-1}
\end{table}

Note that, the $b^1$ axion is always heavier than the overall volume mode and lighter than
small divisor volume mode in large volume limit. This is quite expected because $b^1$ flatness
is expected to get lifted with the standard (unfluxed) $E3$-instanton correction due to an implicit appearance of $b^1$ (and not $c^1$) in the chiral
coordinate $T_\alpha$. The same is reflected through $\Delta_1(\cal F)$ in (\ref{delta}) which has an additional piece $a_s \kappa_{s11}\Theta(0)$, directly coming from $e^{-a_s \, T_s}$ as just have been argued above. This additional piece, in general, causes an imbalance in the fine cancellation of other two terms (we discussed earlier), and results in an order one value. Note that, to nullify this extra pieces $a_s \kappa_{s11}\Theta(0)$ via $\kappa_{s11}=0$ is not sensible as that would mean that small divisor does not have intersection with odd four-cycle and so no odd moduli can be supported on that divisor, and thus things would be too trivialized.

Another important observation in this setup is the fact that there are no tachyons present.
It has been argued in \cite{Conlon:2006tq,Flauger:2009ab} that in a setup equipped with supersymmetric moduli stabilization,
in the presence of flat axionic directions, there are always tachyons. 
However,  such a No-Go theorem does not holds for large volume model in which moduli
stabilization is done in a non-supersymmetric manner \cite{Conlon:2006tq}, and hence there is no conflict in having a flat $\rho_b$-direction and no tachyons.

\subsection*{Case-II : $\kappa_{b11} = 0$}
Let us assume that the big divisor does not intersect with the odd four-cycles, i.e. $\kappa_{b11} =0$.
This is also common when one considers the holomorphic involution which permutes two ``nontrivial identical''
shrinkable del-Pezzo surfaces \cite{Gao:2013pra}. %For shrinkable del-Pezzo divisors, there always
%exist a basis in which only self intersection numbers are non-zero,
%and such surfaces are also called as `diagonal' del-Pezzo.

Using $\kappa_{b11} = 0$ in the ansatz eq.(\ref{eq:Yodd1}), the volume form appearing in the
K\"ahler potential eq.(\ref{eq:Yodd1}) simplifes to \footnote{We thank T. Higaki  for bringing our notice to \cite{Higaki:2011me} where volume form of type \eqref{eq:Yodd2} with $\kappa_{b11}=0$ has been considered (without odd moduli stabilization via F-term scalar potential).} ,
\begin{eqnarray}
\label{eq:Yodd2}
 {\cal Y} = \xi_b (T_b + {\bar T}_b)^{3/2} -\xi_s \left((T_s + {\bar T}_s) +
 \frac{ \kappa_{s11}}{2(S+ {\bar S})}({G}^1+{\bar G}^1)({G}^1+{\bar G}^1) \right)^{3/2}+{\cal C}_{\alpha^\prime} \nonumber\\
\end{eqnarray}
The leading order contributions to the F-term scalar potential can be again collected as three
types of terms, $$ V_{\kappa_{b11}=0} \equiv V_{\alpha'} + V_{np1} + V_{np2} $$
where $ V_{\alpha'}$ and $V_{np1}$ are the same as in eq.(\ref{oddV1}) while $V_{np2}$ is modified,
and is given as under
\bea
\label{OddcaseII}
& & V_{np2} = \frac{\sqrt{2} A_s^2}{3 \,\xi_s \,{\cal V} \sqrt{\tau_s}} \times {\rm exp}
\left[ \frac{a_s  (\kappa_{s11} b^1 b^1-2 g_s \tau_s)}{g_s} + \frac{2  \pi}{g_s}({c^1}^2 g_s^2- {b^1}^2) \right] \\
& & \hskip0.2cm \times \biggl\{ 2 a_s^2 g_s \tau_s |\Theta(b^1,c^1)|^2 +  a_s g_s b^1 ({\rm cos}(a_s  \rho_s)+i \,{\rm sin}(a_s  \rho_s))
\biggl( ( -g_s \pi \bar\Theta'(b^1,c^1) \Theta(b^1,c^1) \nonumber\\
& & \hskip0.4cm +\bar\Theta(b^1,c^1)(-g_s \pi \Theta'(b^1,c^1)+a_s b^1 \kappa_{s11} \Theta(b^1,c^1)+4 g_s \pi b^1 \Theta(b^1,c^1))
{\rm cos}(a_s \rho_s)) \nonumber\\
& & \hskip0.6cm - i (g_s \pi \bar\Theta'(b^1,c^1) \Theta(b^1,c^1) +
\bar\Theta(b^1,c^1)(-g_s \pi \Theta'(b^1,c^1)+a_s b^1 \kappa_{s11} \Theta(b^1,c^1) \nonumber\\
& & \hskip0.8cm -4 i c^1 g_s^2 \pi \Theta(b^1,c^1)){\rm sin}(a_s \rho_s) \biggr)
-\frac{g_s^2 \pi^2}{\kappa_{s11}}(\bar\Theta'(b^1,c^1)-2\bar\Theta(b^1,c^1)(b^1+i \,c^1 g_s)) \nonumber\\
& & \hskip1.2cm ({\rm cos}(2 a_s \rho_s)+ i \, {\rm sin}(2 a_s \rho_s))\biggr\} \nonumber
\eea
where $\Theta'(b^1, c^1) = \theta'_3\left[ -b^1 \pi +i \,c^1 g_s \pi, e^{-g_s \pi} \right]$
and ${\theta_3}'(u,q)$ gives the derivative with respect $u$. It is important to
mention that the simplest critical point which minimizes the potential eq.(\ref{OddcaseII}) is
the same as what was in Case-I eq.(\ref{odd-extreme}). The reason for the same is the fact that
the difference between eq.(\ref{OddcaseII}) and eq.(\ref{oddV1}) effectively vanish at this critical point.
So, it realizes the same LARGE volume non-susy $AdS$ minimum as given by eq.(\ref{eq:AdS}).
\subsubsection*{Axion decay constant and mass matrix}
The axionic sector of the kinetic matrix results in the following estimates for the leading order
contributions to the axion decay constants,
\bea
\label{eq:decayOdd2}
& & f_{\rho_b} \simeq \frac{\sqrt6 \, \xi_b^{2/3}}{{\cal V}^{2/3}} \sim {\cal V}^{-2/3}, \, \, \,
f_{\rho_s} \simeq \frac{{\sqrt{3 \, \xi_s}}}{(2 \tau_s)^{1/4} {\cal V}^{1/2}} \sim {\cal V}^{-1/2} \\
& & f_{b^1} \simeq \frac{\sqrt{3 \, \xi_s  \kappa_{s11} \, \sqrt{2 \tau_s}}}{\sqrt{g_s} \, {\cal V}^{1/2} }\sim {\cal V}^{-1/2}, \, \, \,
f_{c^1} \simeq \frac{\sqrt{3 \, g_s \, \kappa_{s11}\, \xi_s \sqrt{2 \tau_s}}}{{\cal V}^{1/2} } \sim {\cal V}^{-1/2}. \nonumber
\eea
Here, we observe two things; first, the positive definiteness of the kinetic matrix of odd axionic
sector demands that $\kappa_{s11} > 0$ and second, the volume scalings in decay constants for
odd-axions are different from the previous case eq.(\ref{eq:decayOdd}). For the present case,
it has an extra volume suppression of order ${\cal V}^{-1/6}$. %which is good indication for possibility
%of lowering the decay constant values (towards that of QCD axion) using large volume limit.
However, there is a crucial observation that in this case $b^1$-axionic direction is tachyonic. After looking at the eigenvalues of the mass-squared matrix, one gets
the following estimates,
\bea
\label{eq:Mass-odd2}
& & M_{\cal V} \sim \frac{\delta}{{\cal V}^{3/2}}, \,M_{\tau_s} \sim \frac{\delta}{{\cal V}}, \, ; \, \, M_{\rho_b} = 0, \, M_{\rho_s} \sim \frac{\delta}{{\cal V}},\nonumber\\
& & \,  M_{b^1} \sim \frac{\delta\, \Delta_3({\cal F})}{{\cal V}}, \, M_{c^1} \sim \frac{\delta \, \Delta_4({\cal F})}{{\cal V}}
\eea
where $\delta \sim \sqrt{\frac{g_s \, |W_0|^2}{8\pi}}$.
The above volume scaling differs in odd-axionic sector from those of case-I eq.(\ref{eq:Mass-odd1}) while the scaling for the even-moduli sector remain the same.
%Again, the expected results of even-moduli sector are unchanged and are given in the first line. Similar to the previous case-I,
In odd-moduli sector,  $|\Delta_3({\cal F})| \sim {\cal O}(1)$ while $|\Delta_4({\cal F})| < {\cal O}(1)$
for natural model dependent parameters the same as Table \ref{odd1-case-1} except that  $\kappa_{b11}=0$ and $\kappa_{s11}=1$. From Table \ref{odd2case-2}, one realize that $b^1$ modulus direction is always tachyonic
for a generic volume form eq.(\ref{eq:Yodd2}) with $\kappa_{b11} = 0$.
\begin{table}[H]
  \centering
  \begin{tabular}{|c||cc||c|c|cc|}
\hline
  Model  & $\ov{\cal V}$ & $\ov{\tau}_s$ & $m_{\cal V}^2$  &  $m_{\tau_s}^2 \sim m_{\rho_s}^2$ & $m_{b^1}^2$ & $m_{c^1}^2$ \\
    \hline \hline
  S1  & $3.7 \times10^{8}$ & $3.93$ & $3.4 \times 10^{-28}$ & $3.1 \times 10^{-16}$ & $ -8.4 \times 10^{-20}$ & $1.7\times 10^{-20}$ \\
\hline
S2  & $1.0\times10^{8}$ & $3.93$ & $3.3 \times 10^{-28}$ & $3.1 \times 10^{-16}$ & $-3.5 \times 10^{-20}$ & $2.0\times 10^{-34}$ \\
\hline
  \end{tabular}
  \caption{\small Stabilized values of (divisor) volume moduli along with the eigenvalues of mass-squared matrix
  (evaluated at the minimum) in $Mp = 1$ units.
  The respective stabilized values for the axion are $\ov\rho_s = 0 = b^1 = c^1$.}
  \label{odd2case-2}
\end{table}

%Further, on technical grounds, there are more unpleasant issues expected in the present case.
%One possibility to get a volume form of type eq.(\ref{eq:Yodd2}) is simply to consider rigid
%shrinkable four-cycles $D_a$ and $D_b$
%which are mirror to each other under involution.
%The equivariant cohomology of the small divisor $D_s$ under such involution may have non-zero
%$h^{00}_-(D_s) $ component,
%an E3-instanton wrapping on such
% $D_s$ will have non-zero  universal zero modes and therefore result in a $U(1)$
%instanton which does not contribute to the holomorphic superpotential.
%%In such cases, one end up with $h^{00}_- \ne 0$ and hence this would give rise to a $U(1)$ instanton.
%This indicates that the superpotential ansatz we assumed in eq.(\ref{oddW0}) may not be true.
%Although this doesn't exclude the possibility to find a proper example with $\kappa_{b 1 1}=0$ to contribute
%superpotential, it indicates that
% it is better to start with
%(less simple examples with) $\kappa_{b11} \ne 0$ for  model building with odd-axions. 

Note that, in both cases with different volume forms eq.(\ref{eq:Yodd1}) and eq.(\ref{eq:Yodd2}) studied in this section,
the model dependent parameters are chosen such that volume mode avoids the cosmological moduli problem. Further, it is observed that $c^1$ axion is not the lightest for generic values of string coupling. For string coupling $g_s \sim 0.5$, we find that overall volume mode is the lightest, and thus a mass hierarchy is not very generic. For building inflationary model of odd axion $c^1$, one has to consider a small enough string coupling which is well consistent and very natural in the large volume limits. In the next section, we will investigate a poly-instanton LARGE volume setup in which a mass hierarchy (in a part of even and odd sector) is manifestly present via a subdominant poly-instanton correction on top of standard non-peturbative effect. %We look at the insights via moduli stabilization.
%The masses $m_{\cal V} \sim 10$TeV and $\ov{\cal V} \sim, 10^8$ in our sampling does not result in mass of $c^1$
%axion to be in QCD axion mass range $m_{QCD} \sim 10^{-30}\, M_p$! By parametric samplings, one can indeed
%lower $m_{c^1}$ up to the required one, but the same would lower the mass of
%volume mode to suffer the cosmological moduli problem.
%Recall that, in general, the instanton fluxes can help to avoid chirality isseus.
%Hence divisor supporting visible sector may have a superpotnetial contribution.
%But instanton fluxes do not help in lifting the universal right handed zero mode counted by $h^{00}_-$.
%Another good thing for model building is that it does not have a volume mode as a saxion partner.

\section{Extended Poly-Instanton LARGE Volume Scenario}
\label{sec-PolyLVS}
In this section, we start with a short review of the moduli stabilization mechanism after
implementing the poly-instanton corrections in the standard LARGE volume scenario. The hierarchy
of the poly-instanton contribution at the level of superpotential appears in the F-term scalar potential also.
This hierarchial contribution facilitates the moduli stabilization process to be performed in three
steps \cite{Blumenhagen:2012ue}. After stabilizing all the complex structure
moduli and axion-dilaton by the GVW superpotential, the K\"aehler moduli along with respective $C_4$ axions are stabilized in the next two
steps with/without poly-instanton effects. Unlike previous cases studied regarding odd-moduli stabilization,
we expect to have decoupled standard LARGE volume framework from some of the (lighter) moduli.
Recall that in the earlier cases, the stabilization process of all the even/odd moduli was
coupled because the leading contribution for the odd moduli was originated on top of the standard
E3-instanton correction to the superpotential which is responsible for stabilizing `small' divisor
volume mode. Hence, from the point of view of volume scaling, the masses of the odd axions were
found to be larger than that of the overall volume mode. We will investigate if there is some
improvement in this regard with the inclusion of fluxed poly-instanton corrections. The same is expected due
to the appearance of a new hierarchy among standard E3-instanton and the poly-instanton corrections
to the superpotential.

Before analyzing the poly-instanton setup with presence of odd moduli, let us briefly recall the
relevant features of the moduli stabilization mechanism in the standard poly-instanton setup in
LARGE volume scenario. We consider one $C_4$ axion corresponding to the complexified divisor volume
of a del-Pezzo `small' divisor and another $C_4$ axion
complexifying the volume mode of a so-called `Wilson' divisor relevant for generating poly-instanton
contributions to the
superpotential \cite{Blumenhagen:2012Poly1}. The expressions for the K\"{a}hler
potential and the superpotential are,
\bea
\label{eq:KW+race}
 & & K = - 2 \, \ln {\cal Y} \,, \\
 & & W= W_0 + A_s \, e^{-a_s T_s}+ A_s A_w \, e^{-a_s T_s - a_w T_w} \nonumber\\
& & \hskip3cm - B_s \,
e^{-b_s T_s}- B_s B_w \, e^{-b_s T_s - b_w T_w}\,, \nonumber
\eea
where
\eq{
{\cal Y}
= \xi_b (T_b+\bar{T}_b)^{\frac{3}{2}}-\xi_s
(T_s+\bar{T}_s)^{\frac{3}{2}}-\xi_{sw}
\Bigl((T_s+\bar{T}_s) + (T_w+\bar{T}_w)\Bigr)^{\frac{3}{2}} +
C_{\alpha^\prime}\, .
}
Here, we consider a racetrack form of the superpotential as it
has been realized that with a superpotential ansatz without racetrack form, one does not get a minimum
which could be trusted within the regime of validity of Effective field theory description
\cite{Blumenhagen:2012ue}. One can show that the same happens for the present case also. Now, in the large volume limit, (sub)leading contributions to the
generic scalar potential ${\bf V}(\tau_b,\tau_s,\tau_w;\rho_s,\rho_w)$
are simply given as:
\bea
\label{Vgen+racepoly1}
& & {\bf V}({\cal V},\tau_s,\tau_w ;\rho_s, \rho_w) \simeq {\bf V}_{\rm racetrack}^{\rm LVS}({\cal V},\tau_s ;\rho_s)  +
{\bf V}_{\rm poly}({\cal V},\tau_s,\tau_w ;\rho_s, \rho_w)\, \nonumber\\
\eea
As expected, the first term ${\bf V}_{\rm racetrack}^{\rm LVS} \sim {\cal V}^{-3}$ in the
scalar potential eq.(\ref{Vgen+racepoly1}) does not
depend on the Wilson line  divisor volume modulus $\tau_w$ (along with its respective $C_4$ axion $\rho_w$).
So these directions remain flat at leading order and get lifted
via sub-dominant poly-instanton effects ${\bf V}_{\rm poly} \sim {\cal V}^{-3-p}$, where $p$ is a
model dependent parameter. All the K\"ahler moduli are stabilized with this form of potential
eq.(\ref{Vgen+racepoly1}) resulting in a non-susy $AdS$ minimum, and the details for the same can
be found in \cite{Blumenhagen:2012ue,Gao:2013hn}. Further, in large volume limits, the estimates for axionic decay constants are as,
\bea
& & f_{\rho_b} \simeq {\cal V}^{-2/3}, \, f_{\rho_s} \simeq  {\cal V}^{-1/2} \simeq  f_{\rho_w}
\eea
while the various masses scale as,
\bea
M_{\cal V} \sim \frac{\delta}{{\cal V}^{3/2}}, \,M_{\tau_s} \sim \frac{\delta}{{\cal V}}, \,
M_{\tau_w} \sim \frac{\delta}{{\cal V}^{\frac{2+p}{2}}}\, ; \, \,\nonumber \\
M_{\rho_b} = 0, \, M_{\rho_s} \sim \frac{\delta}{{\cal V}}, \,  M_{\rho_w} \sim \frac{\delta}{{\cal V}^{\frac{2+p}{2}}}
\eea
where $\delta \sim \sqrt{\frac{g_s\, |W_0|^2}{8 \pi}}$. Thus, we observe that the masses
of axionic directions lifted
by a non-perturbative term in the superpotential come out to be of the same order as those of
the respective saxion divisor volume modulus appearing in complexified chiral coordinate $T_\alpha = \tau_\alpha + i \rho_\alpha$.
%Therefore, lowering the value of a $C_4$
%axion mass (via choice of model dependent parameters) up to the mass window needed for a QCD axion
%would lower
%the corresponding saxion mass to a value inconsistent with the fifth force constraints. Moreover,
%this would imply a
%cosmological moduli problem also.

\subsection*{Fluxed Poly-Instanton Corrections}
Assuming that the desired mathematical conditions relevant for generating the poly-instanton corrections
can be satisfied, let us consider the following ansatz for the K\"ahler potential and the superpotential
\bea
\label{oddKWpoly}
& &K= -2\, {\rm ln} {\cal Y} = -2\, {\rm ln} \,{\left( \xi_b \Sigma_b^{3/2} - \xi_s \Sigma_s^{3/2} -
\xi_{sw} \Sigma_{sw}^{3/2} + {{\cal C}_{\alpha^\prime}} \right)}, \nonumber\\
& & W = \int_{X} G_3 \wedge \Omega + A_s \, e^{-\, a_s  T_{s}} -B_s \, e^{-\, b_s  T_{s}} +
\bigl(A_s \, A_w \, e^{-\, a_s  T_{s}} \, e^{-\, a_w  T_{w}} \\
& & \hskip1cm  - B_s \, B_w \, e^{-\, b_s  T_{s}} \, e^{-\, b_w  T_{w}}  \bigr) \left( \sqrt{ g_s}
\,\, e^{\,g_s \, \pi\, G^1 G^1} \, \, {\cal \theta}_3\left[ g_s \pi G^1, e^{-g_s \pi} \right] \right). \nonumber
\eea
where
\begin{eqnarray}
\label{eq:Yodd3}
 & & \Sigma_b = T_b + {\bar T}_b + \frac{ \kappa_{b11}}{2(S+ {\bar S})}({G}^1+{\bar G}^1)({G}^1+{\bar G}^1)  \nonumber\\
 & & \Sigma_s = T_s + {\bar T}_s + \frac{ \kappa_{s11}}{2(S+ {\bar S})}({G}^1+{\bar G}^1)({G}^1+{\bar G}^1) \\
 & & \Sigma_{sw} = T_s  + T_w + {\bar T}_s + {\bar T}_w + \frac{ (\kappa_{s11}+\kappa_{w11})}{2(S+ {\bar S})}({G}^1+{\bar G}^1)({G}^1+{\bar G}^1) \nonumber
\end{eqnarray}
Here, we assume that the E3-instanton wrapping the `Wilson' divisor is fluxed and hence correct
the holomorphic pre-factor via a odd-modulus dependent theta function. Further, the E3-instanton
wrapping the `small' divisor is not fluxed.

In large volume limit, the F-term scalar potential with the aforementioned ansatz eq.(\ref{oddKWpoly})
comes out to be in the following form,
\bea
& & \hskip-1in V({{\cal V}, \tau_s, \tau_w; \rho_s, \rho_w , b^1, c^1}) = V_{\rm LVS}^{\rm ex}({{\cal V}, \tau_s; \rho_s , b^1}) \\
& & \hskip1.2in + V_{\rm poly}^{\rm ex}({{\cal V}, \tau_s, \tau_w; \rho_s, \rho_w , b^1, c^1})\nonumber
\eea
where
\begin{itemize}
 \item $ V_{\rm LVS}^{\rm ex}({{\cal V}, \tau_s; \rho_s ; b^1})$ denotes the {\it extended} version
 of LARGE volume potential with the inclusion of  $b^1$ odd axion\footnote{Recall that $c^1$ flatness
 is present in the absence of fluxes turned-on on the instanton divisor.}. This potential scales as
 ${\cal V}^{-3}$ in large volume limit and stabilizes the heavier moduli $\{{\cal V}, \tau_s; \rho_s,  b^1\}$.
\item
The leading corrections to $c^1$ axion along with the Wislon divisor volume mode $\tau_w$
(and its respective axion $\rho_w$) come from the subdominant contributions $V_{\rm poly}^{\rm ex}({{\cal V},
\tau_s, \tau_w; \rho_s, \rho_w ; b^1, c^1})$ which scales as ${\cal V}^{-3-p}$. Here, the parameter $p$
is model dependent and in the absence of instanton-fluxes, $p> 1$ is required for the `Wilson' divisor
volume mode to be the lightest volume modulus \cite{Blumenhagen:2012ue}.
\end{itemize}
The general expressions for  $V_{\rm LVS}^{\rm ex}({{\cal V}, \tau_s; \rho_s ; b^1})$
and $V_{\rm poly}^{\rm ex}({ \tau_w;  \rho_w, c^1} )$ are given in the Appendix \ref{appendixpoly}.
After stabilizing the heavier moduli at their respective minimum, the subleading scalar potential
can be simplified to\footnote{In the absence of odd moduli sector, this potential eq.(\ref{polyc1})
reduces to a two field poly-instanton setup which has been used to study the possibility of realizing
the non-Gaussianities signatures in \cite{Gao:2013hn}.}
\bea
\label{polyc1}
& & \hskip-1cm {\bf V}_{\rm poly}^{\rm ex}(\tau_w, \rho_w, c^1) =  e^{-a_w \tau_w} \, (\lambda_1 +
\lambda_2 \, \tau_w) \\
& &  \times \, e^{-\pi \, g_s \, {c^1}^2} \, \biggl[ (\Theta(c^1) + \ov\Theta(c^1)) \, \cos[a_w \rho_w]
- i (\Theta(c^1) - \ov\Theta(c^1)) \, \sin[a_w \rho_w]  \biggr] \nonumber
\eea
where $\lambda_0, \lambda_1$ and $\lambda_2$ depend on the stabilized values of the heavier moduli
and other model dependent parameters.  The expressions for the same are given by eq.(\ref{eq:Vtau8}) in
Appendix \ref{appendixpoly}. The potential (\ref{polyc1}) has the following minimum,
$$c^1 = 0, \, \, \, \, a_w \ov\rho_w = N \pi,  \, \, \, \, \, \, \, \ov\tau_w = \frac{1}{a_w} -
\frac{\lambda_1}{\lambda_2}$$
Recall that in the absence of racetrack form of the superpotential, the parameters $\lambda_i$'s
are such that one can not have a minimum which could be trusted in the regime of validity of the
effective field theory. However, the presence of racetrack form introduces more parameters in the
picture which facilitates more freedom in model dependent parameters such that  $\lambda_1 > 0$
and $\lambda_2<0$, and $\ov\tau_w >1$ can be easily realized. Further, following the same strategy
for computing the axion decay constants as well as moduli masses, in large volume limit, we find
\bea
& & f_{\rho_b} \simeq {\cal V}^{-2/3}, \, f_{\rho_s} \simeq  {\cal V}^{-1/2} \simeq  f_{\rho_w}, \,\,
f_{b^1} \simeq \frac{1}{\sqrt{g_s} {\cal V}^{1/3}}, \, \, f_{c^1} \simeq \frac{\sqrt{g_s}}{{\cal V}^{1/3}}
\eea
and
\bea
\label{mpoly}
& & M_{\cal V} \sim \frac{\delta}{{\cal V}^{3/2}}, \,  M_{\rho_b} = 0, \, M_{\tau_s} \sim \frac{\delta}{{\cal V}} \sim M_{\rho_s}, \, \, M_{b^1} \sim \frac{\delta}{{\cal V}^{\frac{7}{6}}} \,
; \, \,\nonumber \\
& & \hskip1.5cm M_{\tau_w} \sim \frac{\delta}{{\cal V}^{1+\frac{p}{2}}}\,  \sim M_{\rho_w}, \, \, M_{c^1} \sim \frac{\delta \, \Delta_5({\cal F})}{{\cal V}^{\frac{2}{3}+\frac{p}{2}}}
\eea
where $\delta \sim \sqrt{\frac{g_s \, |W_0|^2}{8\pi}}$. The first line represents the masses
for those moduli which have been stabilized at the leading order in the absence of poly-instanton
corrections, while the second line represents masses for those moduli which have been stabilized by
fluxed poly-instanton corrections. For obvious reasons, the moduli masses in the even sector are same
as in \cite{Blumenhagen:2012ue}. For  justifying the two-step K\"ahler moduli stabilization, we have to choose model
dependent parameters such that $p>1$. Then the lighter ones which remain flat in the absence of
poly-instanton effects get stabilized after including the same, and we observe different
volume scalings for $\tau_w, \rho_w$ and $c^1$ moduli masses. However, similar to the previous case, we again realize the
same suppression factor $\Delta_s \equiv
(2 \theta_3[0, e^{-\pi \, g_s}] + g_s \, \pi \, \theta_3^\prime[0, e^{-\pi \, g_s}])$ appearing inside $\Delta_5({\cal F})$ for $c^1$ axion mass, and for natural model dependent parameters, $\Delta_s \ll {\cal O}(1)$. To get an idea about the numerics, with the following sampling of parameters (similar to the ones used in \cite{Blumenhagen:2012ue,Gao:2013hn}), 
\bea
\label{samplePoly}
& & \xi_b= \frac{1}{36}, \, \xi_s =\frac{1}{6 \sqrt{2}}, \, \xi_{sw} = \frac{1}{6 \sqrt{2}}, C_{\alpha^\prime}= \frac{0.165}{g_s^{3/2}},  \\
& & W_0 = -20,  g_s = 0.12, \, a_s = \frac{2 \pi}{7}, b_s= \frac{2 \pi}{6},\, a_w = 2 \pi = b_w, \nonumber\\
& & A_s = 3, A_w = 0.5, B_s = 2, B_w = 1.749, \kappa_{b11}=-1=\kappa_{s11}. \nonumber
\eea
we have the stabilized values of moduli in the simplest minimum as
\bea
\label{Npolyextremum}
& & \ov{\cal V} \sim 904.86, \,  \ov{\tau}_s \sim 5.68, \, \ov{\tau}_w \sim 1.73, \, \ov{\rho}_s = 0 = \ov{\rho}_w, \, \ov{b^1} = 0 = \ov{c^1} 
\eea
Further, utilizing the \eqref{samplePoly} and \eqref{Npolyextremum}, we have the following estimates for masses of the canonically normalized moduli and the axions (in $M_p = 1$ units)
\bea
\label{Nmasspoly}
& & M_{\cal V} \sim 2.0 \times 10^{-4}, \, M_{\tau_s} \sim 1.8 \times 10^{-2}, \, M_{\tau_w} \sim 2.0 \times 10^{-5} \\
& & M_{\rho_b} =0, \, M_{\rho_s} \sim 1.8 \times 10^{-2}, \, M_{\rho_w} \sim 2.0 \times 10^{-5} \nonumber\\
& & M_{b^1} \sim 2.4 \times 10^{-3}, \, M_{c^1} \sim 3.3 \times 10^{-11}. \nonumber
\eea
which reflects the following mass hierarchies $$M_{\tau_s} \sim M_{\rho_s} > M_{b^1} >  M_{\cal V} > M_{\tau_w} \sim M_{\rho_w} >M_{c^1}$$
as discussed earlier. 
Further, although in the  present case, the risk of $c^1$ axion being heavier than volume mode can be
easily avoided for $p>1$, however,
now the problem with lowering of axion mass along with those of saxion in model dependent 
way reappears through the lowering the mass of `Wilson' divisor volume mode. This can be seen from \eqref{mpoly} that if $\Delta_5({\cal F})$ becomes order one by increasing the string coupling as seen from the plots \ref{Delta_s}, then simple volume scaling shows that $M_{\tau_w} \sim M_{\rho_w} > M_{c^1}$. 
%Finally, before summarizing the analysis done in this paper, let us comment on the mass of the
%axion corresponding to the Big divisor volume mode. In the standard moduli stabilization process
%of LARGE volume scenario, the involutively even $C_4$
%axion corresponding to the K\"ahler coordinate $T_b$ remains flat. It neither appears in K\"ahler
%potential nor in the superpotential. However, if one assumes that a superpotential term of type
%$A_b \, e^{-a_b T_b}$ is allowed via say rigidifying the instanton divisor through magnetic fluxes
%\cite{Bianchi:2011qh}, then the moduli stabilization still results
%in hierarchially separated values of divisor volume moduli, this induces a term in
%the scalar potential of the form\footnote{We neglect the double exponentially suppressed
%term of the form $e^{-2 a_b \tau_b}$ similar to what has been done for poly-instanton case.},
%\bea
%& & e^K \, K^{T_B {\bar T}_i}(D_{T_B} W({T_B})) \, ({\bar D}_{{\bar T}_i} {\bar W}({{\bar T}_i}))+h.c. \sim \frac{e^{- a_b {\cal V}^{2/3}}}{{\cal V}^2}  \subset V  , \, \, \, \, i\ne B
%\eea
%One can simply expect that this would result in a mass for the axion given as
%\bea
%M_{\rho_b} \sim \frac{1}{{\cal V}^{\#}} \, {e^{- \frac{a_b}{2} \times {\cal V}^{2/3}}}
%\eea
%where $\#$ depends on the volume powers appearing via Hessian and inverse K\"ahler metric,
%and a naive scaling estimate implies $\# = 1/3$. For the natural values of model dependent parameters
%and a considerably large volume, the aforementioned mass of $\rho_b$ axion is negligible.

\section{Conclusion}
\label{sec_Conclusions}
In this article, we revisited the F-term moduli stabilization in an extended LARGE volume setup equipped with the
involutively odd moduli. First, we considered a simple extension of large volume setup with the inclusion of a single odd modulus, and investigated the odd axion stabilization with the inclusion of instanton flux effects. Then, we extended the analysis into a poly-instanton LARGE volume framework and revisited the moduli stabilization in the presence of odd moduli. We also computed the masses and decay constants for various even/odd axions present in the respective setups. Subsequently, we realized a mass hierarchy among the divisor volume moduli masses and even/odd axion masses which might be helpful in exploring the inflationary implications of odd-axions. Further, it is also desired to implement this moduli stabilization process in a less simple Type IIB orientifold setup which supports an `explicit' MSSM-like visible sector and subsequently explore the utility of odd axions for studying various cosmo/pheno aspects in the regime of Axion Physics.

\section*{Acknowledgement}
We are grateful to Ralph Blumenhagen for several enlightening clarifications, discussions and suggestions. We also thank Thomas Grimm, Sebastian Halter for useful discussions, and Arthur Hebecker, Savdeep Sethi for useful comments. XG
is supported by the MPG-CAS Joint Doctoral Promotion Programme and PS is supported
by a postdoctoral research fellowship from the Alexander von Humboldt Foundation.

\renewcommand{\theequation}{A\arabic{equation}}
  % redefine the command that creates the equation no.
  \setcounter{equation}{0}  % reset counter

\newpage
\appendix
\section{Kinetic matrices for various $K$ ansatz}
\label{appendixOdd}
All the kinetic terms for the respective moduli in a given ansatz for the K\"ahler potential can be  written out utilizing the K\"ahler metric as below,
\eq{
{\cal L}_{\rm kinetic}({\cal V},\tau_s,...; \rho_b,\rho_s,...;b^1, c^1, ...) \equiv
K_{I\bar J} (D_{\mu} T_I) ({\bar D}^\mu {\bar T}_{\bar J}) \,,}
where the K\"ahler potential is generically defined as $K \equiv -2 \ln {\cal Y} = -2 \ln({\cal V}+C_{\alpha^\prime})$, ${\cal V}$ being overall volume of the Calabi Yau. Neglecting the $\alpha^\prime$-corrections $C_{\alpha^\prime}$, in this section, we present the kinetic matrix for each of the K\"ahler potential ansatz studied in this article. The kinetic matrix when evaluated at the respective minimum of the potential is found to be of block diagonal form with three blocks corresponding to divisor volume moduli, respective $C_4$ axions and odd ($B_2, C_2$) axions, if present in the ansatz for $K$. The relevant volume scalings in the even/odd axion decay constants can be easily  estimated utilizing large volume limit.
%\subsection{Ansatz: $${\cal V} = \xi_b (T_b+\bar{T}_b)^{\frac{3}{2}}-\sum_{s =2}^{h^{1,1}_+} \, \xi_s (T_s+\bar{T}_s)^{\frac{3}{2}}$$}
\subsection*{Standard LARGE volume setup}
For the volume form of type
$${\cal V} = \xi_b (T_b+\bar{T}_b)^{\frac{3}{2}}-\sum_{s =2}^{h^{1,1}_+} \, \xi_s (T_s+\bar{T}_s)^{\frac{3}{2}},$$ the non-zero components of the kinetic matrix are as under
\bea
& & K_{{\cal V}{\cal V}} =  \frac{1}{3{\cal V}^2},  \, \, K_{{\cal V}{\tau_s}}=-\frac{3\, \xi_s \, {\sqrt \tau_s}}{\sqrt2 \, {\cal V}^2}=K_{{\tau_s}{\cal V}}, \\
& & K_{{\tau_s}{\tau_s}}=\frac{3 \,\xi_s }{2\sqrt2 \, {\sqrt\tau_s} \, {\cal V}}, \, \, K_{{\tau_s}{\tau_r}}=-\frac{3 \xi_s\, \xi_r\,{\sqrt{\tau_s\tau_r}}}{{\cal V}^2} =  K_{{\tau_r}{\tau_s}}, \nonumber\\
& & K_{{\rho_b}{\rho_b}} =  \frac{3 \, \xi_b^{4/3}}{{\cal V}^{4/3}}, \, \, K_{{\rho_b}{\rho_s}}=  -\frac{9 \xi_b^{2/3} \, \xi_s \, {\sqrt \tau_s}}{\sqrt2 \,{\cal V}^{5/3}}
=K_{{\rho_s}{\rho_b}},\nonumber\\
& & K_{{\rho_s}{\rho_s}}= \frac{3 \,\xi_s }{2\sqrt2 \, {\sqrt \tau_s} \, {\cal V}}, \, \, K_{{\rho_s}{\rho_r}}= \frac{9 \xi_s \, \xi_r \,{\sqrt{\tau_s\tau_r}}}{{\cal V}^2}
=K_{{\rho_r}{\rho_s}}.\nonumber
\eea
where $\{s, r\} \in \{2, ..., h^{11}_+\} \, \, {\rm and } \, \, s \neq r $ is considered in cross terms.
\subsection*{Extended LARGE volume setup}
For the volume form of type
\begin{eqnarray}
 & & \hskip-1cm {\cal V} =  \xi_b \, \left( (T_b + {\bar T}_b) + \frac{ \kappa_{b11}}{2(S+ {\bar S})}({G}^1+{\bar G}^1)({G}^1+{\bar G}^1)\right)^{\frac{3}{2}}  \\
& & \hskip1in - \xi_s \, \left((T_s + {\bar T}_s) + \frac{ \kappa_{s11}}{2(S+ {\bar S})}({G}^1+{\bar G}^1)({G}^1+{\bar G}^1)\right)^{\frac{3}{2}}, \nonumber
\end{eqnarray}
the non-zero components of the kinetic matrix are as under
\bea
& & K_{{\cal V}{\cal V}} =  \frac{1}{3{\cal V}^2},  \, \, K_{{\cal V}{\tau_s}}=-\frac{3\, \xi_s \, {\sqrt \tau_s}}{\sqrt2 \, {\cal V}^2}, \, \, K_{{\cal V} b^1} = \frac{\xi_b^{2/3} \, \kappa_{b11} b^1}{g_s \, {\cal V}^{5/3}} = -\frac{1}{g_s} K_{{\cal V} c^1}\nonumber\\
& & K_{{\tau_s}{\tau_s}}=\frac{3 \,\xi_s }{2\sqrt2 \, {\sqrt\tau_s} \, {\cal V}}, \, \, K_{{\tau_s} b^1} = \frac{3 \, \xi_s  \, \kappa_{s11} b^1}{2 \sqrt{2} \, g_s \,\sqrt{\tau_s} \, {\cal V}} = - \frac{1}{g_s} \, K_{{\tau_s} c^1},  \nonumber\\
& & K_{b^1 b^1} = -\frac{3\, \kappa_{b11}\, \xi_b^{2/3}}{2 \, g_s \,{\cal V}^{2/3}} = \frac{1}{g_s^2} \, K_{c^1 c^1}, \, \, K_{{\rho_b}{\rho_b}} =  \frac{3 \, \xi_b^{4/3}}{{\cal V}^{4/3}}, \\
& & K_{{\rho_b}{\rho_s}}=  -\frac{9 \xi_b^{2/3} \, \xi_s \, {\sqrt \tau_s}}{\sqrt2 \,{\cal V}^{5/3}}, \, \, K_{{\rho_b}{b^1}}= \frac{3 \, \xi_b^{4/3} \, \kappa_{b11} b^1}{g_s{\cal V}^{4/3}} = -\frac{1}{g_s} K_{{\rho_b}{c^1}}\nonumber\\
& & K_{{\rho_s}{\rho_s}}= \frac{3 \,\xi_s }{2\sqrt2 \, {\sqrt \tau_s} \, {\cal V}}, \, \, K_{{\rho_s}{b^1}}= \frac{3 \, \xi_s  \, \kappa_{s11} b^1}{2 \sqrt{2} \, g_s \,\sqrt{\tau_s} \, {\cal V}} = - \frac{1}{g_s} \, K_{{\rho_s} c^1}.\nonumber
\eea
Therefore, the Kinetic matrix is block diagonal in even/odd sector only when evaluated at the minimum which requires $b^1 = 0$.
\subsection*{Poly-instanton setup}
For the volume form of type
$${\cal V} = \xi_b (T_b+\bar{T}_b)^{\frac{3}{2}}-\xi_s
(T_s+\bar{T}_s)^{\frac{3}{2}}-\xi_{sw}
\Bigl((T_s+\bar{T}_s) + (T_w+\bar{T}_w)\Bigr)^{\frac{3}{2}},$$
the non-zero independent components of the kinetic matrix are as under
\bea
\label{eq:KEpoly}
& &  K_{{\cal V}{\cal V}} =  \frac{1}{3 {\cal V}^2} ,  \, \, K_{{\cal V}{\tau_s}}= -\frac{3 (\sqrt{{\tau_s}} \xi_s+\xi_{sw} \sqrt{{\tau_s}+{\tau_w}})}{\sqrt{2}
   {\cal V}^2},  \nonumber\\
& & K_{{\cal V}{\tau_w}}=-\frac{3\, \xi_{sw} \sqrt{{\tau_s}+{\tau_w}}}{\sqrt{2} {\cal V}^2}, \, \, K_{{\tau_s}{\tau_s}}=\frac{3
   \left(\frac{\xi_s}{\sqrt{{\tau_s}}}+\frac{\xi_{sw}}{\sqrt{{\tau_s}+{\tau_w}}}\right)}{2 \sqrt{2} {\cal V}},\\
& & K_{{\tau_s}{\tau_w}}=-\frac{3 \xi_{saw}}{2 \sqrt{2} {\cal V} \sqrt{{\tau_s}+{\tau_w}}}, \, \, K_{{\tau_w}{\tau_w}} = \frac{3 \, \xi_{sw}}{2 \sqrt{2} {\cal V} \sqrt{{\tau_s}+{\tau_w}}}, \nonumber\\
& & K_{{\rho_b}{\rho_b}} = 3 \left(\frac{\xi_b}{{\cal V}}\right)^{4/3}, \, \, K_{{\rho_b}{\rho_s}}=   -\frac{9 \, \xi_b^{2/3} \left(\sqrt{{\tau_s}} \xi_s + \xi_{sw}
   \sqrt{{\tau_s}+{\tau_w}}\right)}{\sqrt{2} {\cal V}^{5/3}},\nonumber\\
& &  K_{{\rho_b}{\rho_w}}=   -\frac{9 \, \xi_b^{2/3} \xi_{sw} \sqrt{{\tau_s}+{\tau_w}}}{\sqrt{2} {\cal V}^{5/3}}, \, \, K_{{\rho_s}{\rho_s}}=  \frac{3 \left(\frac{\xi_s}{\sqrt{{\tau_s}}}+\frac{\xi_{sw}}{\sqrt{{\tau_s}+{\tau_w}}}\right)}{2 \sqrt{2} {\cal V}}, \nonumber\\
& &  K_{{\rho_s}{\rho_w}}= \frac{3 \xi_{sw}}{2 \sqrt{2} {\cal V} \sqrt{{\tau_s}+{\tau_w}}}, \, \,  K_{{\rho_w}{\rho_w}}= \frac{3 \xi_{sw}}{2 \sqrt{2} {\cal V} \sqrt{{\tau_s}+{\tau_w}}}. \nonumber
\eea
\subsection*{Extended poly-instanton setup}
For the volume for os type
$${\cal V} = \xi_b \Sigma_b^{3/2} - \xi_s \Sigma_s^{3/2} - \xi_{sw} \Sigma_{sw}^{3/2},$$
where
\begin{eqnarray}
 & & \Sigma_b = T_b + {\bar T}_b + \frac{ \kappa_{b11}}{2(S+ {\bar S})}({G}^1+{\bar G}^1)({G}^1+{\bar G}^1)  \nonumber\\
 & & \Sigma_s = T_s + {\bar T}_s + \frac{ \kappa_{s11}}{2(S+ {\bar S})}({G}^1+{\bar G}^1)({G}^1+{\bar G}^1) \nonumber\\
 & & \Sigma_{sw} = T_s  + T_w + {\bar T}_s + {\bar T}_w + \frac{ (\kappa_{s11}+\kappa_{w11})}{2(S+ {\bar S})}({G}^1+{\bar G}^1)({G}^1+{\bar G}^1) \nonumber
\end{eqnarray}
the non-zero independent components of the kinetic matrix are as in eq.(\ref{eq:KEpoly}) along with the following extra components in odd sector,
\bea
\label{eq:KEpolyE}
& &  K_{{\cal V}\, b^1} = \frac{\xi_b^{2/3} \, \kappa_{b11} b^1}{g_s \, {\cal V}^{5/3}} = -\frac{1}{g_s} K_{{\cal V}\, c^1} ,  \nonumber\\
& & K_{\tau_s b^1}=\frac{3 \, b^1 \left(\frac{\xi_s \, \kappa_{s11}}{\sqrt{\tau_s}} +\frac{\xi_{sw} \, (\kappa_{s11} +\kappa_{w11}) }{\sqrt{\tau_s+\tau_w}}\right)}{2\sqrt{2} \,g_s \, {\cal V}} = -\frac{1}{g_s} K_{\tau_s c^1},\\
& & K_{\tau_w b^1}=\frac{3 \, \xi_{sw}  \, (\kappa_{s11} + \kappa_{s11}) b^1}{2 \sqrt{2} \, g_s \,\sqrt{\tau_s+\tau_w} \, {\cal V}} = -\frac{1}{g_s} K_{\tau_w c^1}, \nonumber\\
& & K_{b^1 b^1}=- \frac{3  \, \kappa_{b11} \xi_b^{2/3}}{2 \, g_s \, {\cal V}^{2/3}} = \frac{1}{g_s^2} K_{c^1 c^1}, \nonumber\\
& & K_{{\rho_b}{b^1}}= \frac{3 \, \xi_b^{4/3} \, \kappa_{b11} b^1}{g_s{\cal V}^{4/3}} = -\frac{1}{g_s} K_{{\rho_b}{c^1}}\nonumber\\
& & K_{{\rho_s}{b^1}}= \frac{3 \, b^1 \left(\frac{\xi_s \, \kappa_{s11}}{\sqrt{\tau_s}} +\frac{\xi_{sw} \, (\kappa_{s11} +\kappa_{w11}) }{\sqrt{\tau_s+\tau_w}}\right)}{2\sqrt{2} \,g_s \, {\cal V}} = -\frac{1}{g_s} K_{\tau_s c^1}, \nonumber\\
& & K_{\rho_w b^1}=\frac{3 \, \xi_{sw}  \, (\kappa_{s11} + \kappa_{s11}) b^1}{2 \sqrt{2} \, g_s \,\sqrt{\tau_s+\tau_w} \, {\cal V}} = -\frac{1}{g_s} K_{\rho_w c^1}. \nonumber
\eea

\section{Scalar potential and Moduli Stabilization}
\label{appendixpoly}
In the large volume limit, (sub)leading contributions to the
generic scalar potential ${\bf V}({\cal V},\tau_s,\tau_w;\rho_s,\rho_w, b^1, c^1)$
are simply given as:
\bea
\label{eq:Vgen+racepoly}
& & \hskip-2cm {\bf V}({\cal V},\tau_s,\tau_w ;\rho_s, \rho_w, b^1, c^1) \simeq {\bf V}_{\rm LVS}^{\rm ex}
({\cal V},\tau_s ;\rho_s, b^1) \\
& & \hskip1.4in +{\bf V}_{\rm poly}^{\rm ex}({\cal V},\tau_s,\tau_w ;\rho_s, \rho_w, b^1, c^1)\, \nonumber
\eea
In eq.(\ref{eq:Vgen+racepoly}), the leading contributions for the two type of terms are given as under
\bea
\label{eq:Vgen+racepoly1}
& & \hskip -2cm
{\bf V}_{\rm LVS}^{\rm ex}({\cal V},\tau_s ;\rho_s, b^1)  =  \frac{3 {\, {\cal C}_{\alpha^\prime}} \,|W_0|^2}{2 \, {\cal V}^3} +
\frac{2\sqrt2\, a_s^2 \, \mu_1^2 \, \sqrt{{\tau_s}}}{3\,  \xi_s \, {\cal V}}
+\frac{2 \sqrt2\, b_s^2 \,\mu_2^2 \, \sqrt{{\tau_s}}}{3\, \xi_s \, {\cal V}}\\
& & \hskip 2cm
 + \frac{4\, a_s \, W_0\, \mu_1 \,  {\tau_s} \cos[\, a_s {\rho_s}]}{{\cal V}^2}
 -\frac{4\, b_s \, W_0\, \mu_2 \,  {\tau_s} \cos [\, b_s {\rho_s}]}{{\cal V}^2} \nonumber\\
& &  \hskip 3cm
-\frac{4 \sqrt2 \,a_s \, b_s \, \mu_1 \, \mu_2 \, \sqrt{{\tau_s}} \cos \bigl[(\, a_s-\, b_s){\rho_s}\bigr]}{3\, \xi_s \,{\cal V}}
  \nonumber
\eea
where $\mu_1 = A_s \, e^{- \,a_s \left({\tau_s} -\frac{\kappa_{s11} \, b^1 b^1}{2\, g_s} \right)}$ and $\mu_2 = B_s \, e^{- \,b_s \left({\tau_s} -\frac{\kappa_{s11} \, b^1 b^1}{2\, g_s} \right)}$.
The extrema  of the above leading order scalar potential
can be  collectively described  by the intersection of the following hypersurfaces in moduli space
\bea
\label{hyperextr}
& &  b^1 = 0 , \, a_s \ov\rho_s = N \pi \, , \, \, {\rm where } \, \, \, N \in {\mathbb Z} \, ;\nonumber\\
& & W_0 \simeq
\frac{\ov{\cal V} \,(b_s \ov\mu_2 -a_s \ov\mu_1) \, \Bigl[b_s \ov\mu_2 (-1 + 4 b_s \ov\tau_s)-a_s \ov\mu_1 (-1 + 4 a_s \ov \tau_s)\Bigr]}{6 \sqrt2 \, \xi_s \, \sqrt{\ov\tau_s} \Bigl[b_s \ov\mu_2 (-1 + b_s \ov\tau_s)-a_s \ov\mu_1 (-1 + a_s \ov\tau_s)\Bigr]}\,;\\
& & {\cal C}_{\alpha^\prime} \simeq \frac{32 \sqrt{2} \, \xi_s \, \ov\tau_s^{\frac{5}{2}} (b_s^2\, \ov\mu_2 -
a_s^2\, \ov\mu_1) \, \Bigl[b_s \ov\mu_2 (-1 + b_s \ov\tau_s)-a_s \ov\mu_1 (-1 + a_s \ov\tau_s)\Bigr]}{\Bigl[ a_s \ov\mu_1(-1 + 4 a_s \ov\tau_s)
- b_s \ov\mu_2(-1 + 4 b_s \ov\tau_s) \Bigr]^2} \, ;\nonumber\\
& &   {\rm with} \quad
\ov\mu_1 \equiv \mu_1(b^1 = 0)= A_s\, e^{-a_s \ov\tau_s} \quad \text{~and~} \ \  \ov\mu_2 \equiv \mu_2(b^1 = 0)= B_s\, e^{-b_s
  \ov\tau_s}. \nonumber
\eea
After stabilizing the (heavier) moduli $\{{\cal V}, \tau_s, \rho_s, b^1\}$ via the aforementioned extermination conditions eq.(\ref{hyperextr}), the second part of the expression eq.(\ref{eq:Vgen+racepoly}) which is subleading contribution coming from the poly-instanton corrections simplifies to the form below
\bea
& & \hskip-1cm {\bf V}_{\rm poly}^{\rm ex}(\tau_w, \rho_w, c^1) =  e^{-a_w \tau_w} \, (\lambda_1 + \lambda_2 \, \tau_w) \\
& &  \times \, e^{-\pi \, g_s \, {c^1}^2} \, \biggl[ (\Theta(c^1) + \ov\Theta(c^1)) \, \cos[a_w \rho_w] - i (\Theta(c^1) - \ov\Theta(c^1)) \, \sin[a_w \rho_w]  \biggr] \nonumber
\eea
where
\eq{
\label{eq:Vtau8}
\lambda_1 &= \lambda_0 \biggl[4 \ov\tau_s \bigl( (a_s -a_w) A_w \ov\mu_1 -(b_s -a_w) B_w \ov\mu_2\bigr)\\
 &+\frac{\ov\tau_s\, (b_s B_w \ov\mu_2-a_s A_w \ov\mu_1)
\,\bigl(a_s \ov\mu_1 (-1+4a_s \ov\tau_s)-b_s \ov\mu_2 (-1+4b_s
  \ov\tau_s)\bigr)
}{a_s \ov\mu_1 (-1+a_s \ov\tau_s)-b_s \ov\mu_2 (-1+b_s \ov\tau_s)}
\biggr]\,,\\[0.2cm]
\lambda_2 &= \lambda_0 a_w \left[\frac{\bigl( B_w \ov\mu_2- A_w \ov\mu_1\bigr) \bigl(a_s \ov\mu_1 (-1+4a_s \ov\tau_s)
-b_s \ov\mu_2 (-1+4b_s \ov\tau_s)\bigr)}{a_s \ov\mu_1 (-1+a_s
    \ov\tau_s)-b_s \ov\mu_2 (-1+b_s \ov\tau_s)}\right]}
with
\eq{
\lambda_0 = \frac{\sqrt{g_s} (\, a_s \ov\mu_1-b_s \ov\mu_2)}{3 \,\sqrt{2} \, \xi_s
  \,\ov{\cal V} \, \sqrt{\ov\tau_s}}\, . \nonumber}

\section{Expressions for odd-axion masses}
\label{appendixMass}
\begin{itemize}
\item
The expressions for squared-mass values of the odd-moduli for $\kappa_{b11} \neq 0$ case are given as,
\bea
\label{oddMass1}
& & M_{b^1,b^1}=\frac{48 \sqrt{2} \xi_s W_0^2 \tau_s^{3/2} (-1+a_s \tau_s)(g_s \pi^2 \Theta''(0)
+ 2 \pi \Theta(0) +a_s \kappa_{s11}\Theta(0))}{ {\xi_b}^{2/3} \kappa_{b11} \Theta(0) {\cal V}^{7/3}
(1-4 a_s \tau_s)^2 }\nonumber\\
& & M_{c^1,c^1}=-\frac{48 \sqrt{2} \xi_s W_0^2 \tau_s^{3/2} (-1+a_s \tau_s)(g_s \pi^2 \Theta''(0)
+2 \pi \Theta(0) ) }{ {\xi_b}^{2/3} \kappa_{b11} \Theta(0) {\cal V}^{7/3} (1-4 a_s \tau_s)^2 }
\eea
\item
For $ \kappa_{b11} = 0 $, the squared-mass values of odd-moduli become:
\bea
& & M_{b^1,b^1}=-\frac{4 W_0^2(-1+a_s \tau_s)}{a_s^2 \kappa_{s11}^2 {\cal V}^2 \Theta(0)^2 (1-4 a_s \tau_s)}
\times \biggl\{-2 g_s^2 \pi^2 (\pi \Theta''(0)  +2 \Theta(0))^2  \nonumber\\
& & \hskip1cm
 \times (-1+a_s \tau_s) +a_s g_s \kappa_{s11} \pi \Theta(0)\biggl(\pi \Theta'(0)(4-a_s \tau_s)-8 \Theta(0) \nonumber\\
& & \hskip1cm \times (-1+a_s \tau_s)\biggr) + a_s^2 \kappa_{s11} \Theta(0)^2 (6 \pi \tau_s+\kappa_{s11}(2+a_s \tau_s))  \biggr\} \\
& & M_{c^1,c^1}=\frac{4 \pi W_0^2(-1+a_s \tau_s)}{a_s^2 \kappa_{s11}^2 {\cal V}^2 \Theta(0)^2 (1-4 a_s \tau_s)}
\times \biggl\{3 a_s^2 g_s \tau_s \kappa_{s11} \pi \Theta''(0) \Theta(0) \nonumber\\
& & \hskip1cm
+ 6 a_s^2 \tau_s \kappa_{s11} \Theta(0)^2 + 2 g_s \pi (\pi \, g_s \, \Theta''(0)+2 \Theta(0))^2(-1+a_s \tau_s) \biggr\} \nonumber
\eea
\item
The squared-mass expression for odd moduli in extended poly-instanton setup are as under,
\bea
& & M_{b^1,b^1}= \frac{48 \sqrt{2} \, \xi_s \, \kappa_{s11} \, (b_s^2 \ov\mu_2 -a_s^2 \ov\mu_1) \, |W_0|^2 \, \ov\tau_s^{3/2}}{\xi_b^{2/3} \, \kappa_{b11} \, \ov{\cal V}^{7/3} \, (a_s \, \ov\mu_1(-1+4 a_s \ov\tau_s)-b_s \, \ov\mu_2(-1+4 b_s \ov\tau_s))^2}\nonumber\\
& & \hskip1.5cm \times \, \left(-a_s \ov\mu_1 (-1 + a_s \ov\tau_s) + b_s \ov\mu_2 (-1 + b_s \ov\tau_s)\right) \nonumber\\
& & M_{c^1,c^1}=\frac{g_s \, \lambda_2 \, e^{-1+\frac{a_w \lambda_1}{\lambda_2}} \, \left(2 \Theta(0) + \pi \, g_s \, \Theta''(0) \right)}{3 \, a_w \, \kappa_{b11}}
\eea
where $\lambda_i$s and $\ov\mu_i$s are defined in previous sections \ref{appendixpoly}.
\end{itemize}
In all above expressions, $\Theta'(0)=\theta_3'[-b^1 \pi +i \, c^1 g_s \pi, e^{-g_s \pi}]|_{b^1=0,c^1=0}$. The explicit expressions for masses of even sector moduli can be found in earlier work \cite{Gao:2013hn}.

\clearpage
\nocite{*}
\bibliography{OddQCDAxion}
\bibliographystyle{utphys}

\end{document}